\documentclass[conference]{IEEEtran}
\usepackage[pdftex]{graphicx}
\usepackage{cite}
\usepackage{amsfonts}
\usepackage{algorithm}
\usepackage{algorithmicx}
\usepackage{algpseudocode}
\usepackage{subfigure}
\usepackage{amsfonts}
\usepackage{amsmath,amssymb}
\usepackage{bm}
\usepackage{textcomp}

\algnewcommand\algorithmicforeach{\textbf{for each}}
\algdef{S}[FOR]{ForEach}[1]{\algorithmicforeach\ #1\ \algorithmicdo}

\algdef{SE}[DOWHILE]{Do}{doWhile}{\algorithmicdo}[1]{\algorithmicwhile\ #1}%

\newcommand{\argmax}{\mathop{\rm arg~max}\limits}

\renewcommand{\baselinestretch}{1.}

\ifCLASSINFOpdf
\else
\fi

\begin{document}
%
\title{Wafer-level Variation Modeling for Multi-site RF IC Testing via
  Hierarchical Gaussian Process}

\author{
  \IEEEauthorblockN{Michihiro Shintani, Riaz-Ul-Haque Mian, \\ and Michiko Inoue}
  \IEEEauthorblockA{
    Graduate School of Science and Technology \\
    Nara Institute of Science and Technology\\
    8916-5 Takayama-cho, Ikoma 630-0192, Japan\\
    Email: \{shintani\}@is.naist.jp}
  \and
  \IEEEauthorblockN{Tomoki Nakamura, Masuo Kajiyama, \\and Makoto Eiki}
  \IEEEauthorblockA{Sony Semiconductor Manufacturing Corporation\\
    Nagasaki TEC\\
    1883-43 Tsukuba-machi, Isahaya-shi, \\Nagasaki 854-0065, Japan\\
    Email: \{Tomoki.Nakamura,Masuo.Kajiyama,\\Makoto.Eiki\}@sony.com}
}

\maketitle

\begin{abstract}
  Wafer-level performance prediction has been attracting attention to
  reduce measurement costs without compromising test quality in
  production tests. Although several efficient methods have been
  proposed, the site-to-site variation, which is often observed in
  multi-site testing for radio frequency circuits, has not yet been
  sufficiently addressed. In this paper, we propose a wafer-level
  performance prediction method for multi-site testing that can
  consider the site-to-site variation. The proposed method is based on
  the Gaussian process, which is widely used for wafer-level spatial
  correlation modeling, improving the prediction accuracy by extending
  hierarchical modeling to exploit the test site information provided
  by test engineers. In addition, we propose an active test-site
  sampling method to maximize measurement cost reduction. Through
  experiments using industrial production test data, we demonstrate
  that the proposed method can reduce the estimation error to $1/19$
  of that obtained using a conventional method. Moreover, we
  demonstrate that the proposed sampling method can reduce the number
  of the measurements by 97\% while achieving sufficient estimation
  accuracy.
\end{abstract}





%
\IEEEpeerreviewmaketitle

\section{Introduction}\label{sec:introduction}
Large-scale integrated circuits (LSIs) are now embedded in every
product to support the smooth functioning of our daily lives. In
addition to automobiles, healthcare, and aerospace, which are directly
related to human life, the LSIs are utilized in social infrastructure
that supports our daily lives, such as computer networks, power
transmission systems, and transportation control systems. However,
with the spread of the LSIs, their reliability has become a crucial
issue, and faulty LSIs that do not operate properly not only interrupt
the services of the systems that include them but also lead to a
serious impact on our society.

To guarantee the LSI reliability, multiple test items are tested
and/or measured under various conditions during several stages of LSI
manufacturing. With the increase in scale and multi-functionality of
the LSIs, an increasing number of items need to be tested, leading to
test cost inflation. Thus, it has become a serious problem because the
test cost accounts for most of the LSI manufacturing cost.

Various test cost reduction methods have been proposed that apply data
analytics, machine learning algorithms, and statistical
methods~\cite{TCAD2017_Wang,ETS2018_Stratigopoulos,ITC2018_Shintani}. In
particular, the wafer-level characteristic modeling method based on a
statistical algorithm is the most promising candidate that reduces the
test cost, that is, measurement cost, without impairing the test
quality~\cite{TSM2010_Reda,ICCAD2009_Li,DAC2010_Zhang,TCAD2011_Zhang,DATE2014_Zhang,ICCAD2012_Kupp,ITC2014_Ahmadi,DATE2013_Huang}.
In these studies, a statistical modeling technique was used to predict
the entire measurement on a wafer from a small number of sample
measurements.  Because the estimation eliminates the need for
measurement, it not only reduces the cost of measurement but also can
be used to reduce the number of test items and/or change the test
limits, which is expected to improve the efficiency of adaptive
testing~\cite{DATE2010_Marinissen,ITC2011_Gotkhindikar,ETS2012_Yilmaz,TCAD2014_Shintani}.
In~\cite{TSM2010_Reda}, the expectation-maximization (EM)
algorithm~\cite{EM} was used to predict the
measurement. In~\cite{ICCAD2009_Li,DAC2010_Zhang,TCAD2011_Zhang,DATE2014_Zhang},
a statistical prediction method, called a {\it virtual prove}, based
on compressed sensing~\cite{TIT2006_Donoho,CS} was proposed.  The {\it
  Gaussian process} (GP)-based method~\cite{gp_book} provides more
accurate prediction
results~\cite{ICCAD2012_Kupp,DATE2013_Huang,ITC2014_Ahmadi}.  The use
of GP modeling has another side benefit.  As it calculates the
confidence of a prediction, the user can confirm whether the number
and location of measurement samples are sufficient, which is a
significant advantage from a practical viewpoint.

\begin{figure}[!t]
  \centering
  \subfigure[First lot]{
    \includegraphics[width=0.78\linewidth]{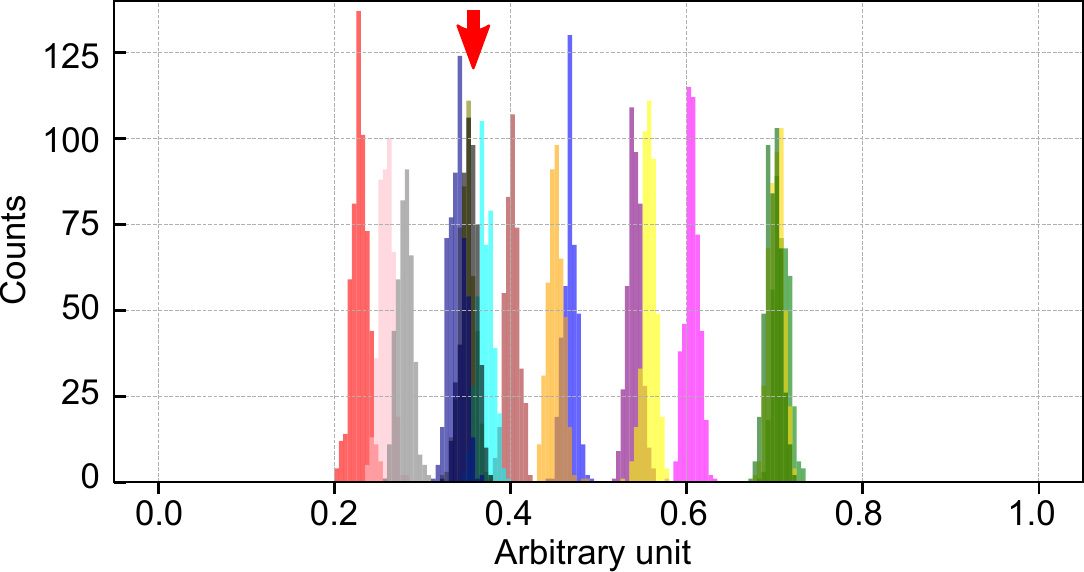}
    \label{fig:hist03}
  }
  \subfigure[Sixth lot]{
    \includegraphics[width=0.78\linewidth]{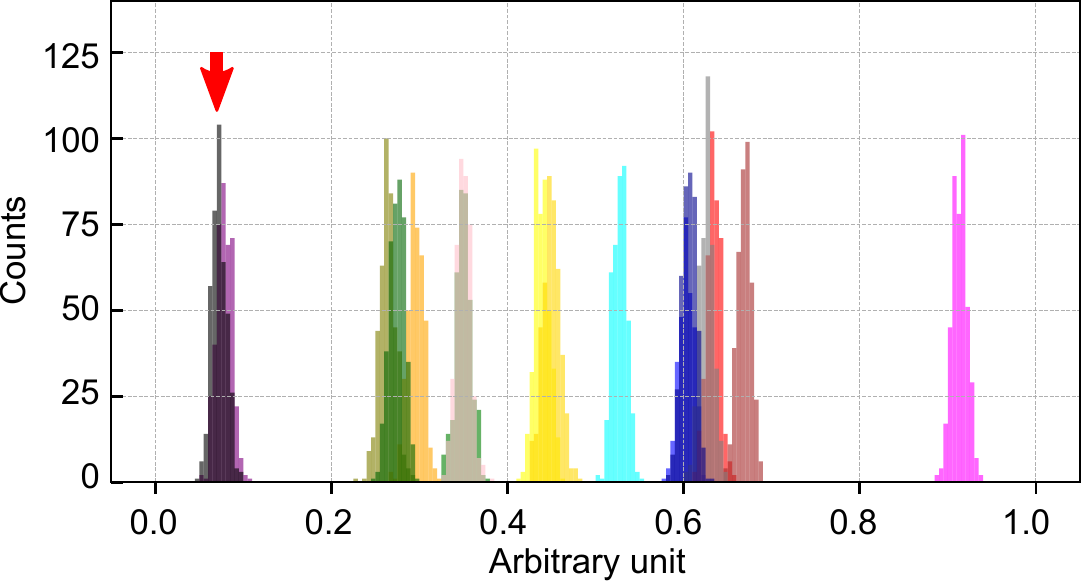}
    \label{fig:hist07}
  }
  \caption{Histograms of measured characteristics of industrial RF
    circuit on a wafer, measured by multi-site testing with 16
    sites. The histograms of each site are shown with different
    colors, i.e., 16 histograms are presented here. We can observe the
    significant variations in the histogram between
    sites. Furthermore, the locations of the black histograms are much
    different between the early and latest lots. The horizontal axis
    is expressed in arbitrary unit.}
  \label{fig:hist}
\end{figure}

Most of these methods assume that the device characteristics on the
wafer gradually change with wafer coordinates; however, this
assumption does not hold for the measurement of radio frequency (RF)
circuits under multi-site
testing~\cite{ITC2012_Sumikawa,ETS2014_Lehner,VTS2014_Farayola}, in
which a probe card is adapted to simultaneously probe multiple devices
under test (DUTs). The contact of the probe card with the DUTs to be
tested is called a {\it touchdown}. Moreover, the position of the
needles in a touchdown is called a {\it site}. During the measurement
of the RF circuit, a calibration circuit for impedance matching is
added on the probe card, causing a much larger variation than the
spatial variation due to its parasitic components, as shown in
Fig.~\ref{fig:hist}.

Figure~\ref{fig:hist03} shows the histograms of the characteristics of
an industrial RF circuits, which are fabricated using a 28\,nm process
technology, measured by a multi-site test with 16 sites per
measurement in the first fabrication lot.  The histograms are shown in
different colors for each site. While low variance can be seen for
each histogram, it is clear that there are significant differences
between the histograms, that is, differences in sites. Most of the
existing methods fail to model this measurement result because of the
discontinuous change between the sites.

Only the work in~\cite{DATE2013_Huang} attempted to solve this issue
of the discontinuous change.  In~\cite{DATE2013_Huang}, a two-step
modeling method using $k$-means clustering~\cite{BJMSP2006_Steinley}
and the GP was proposed.  In the first step, all dies on a wafer are
explicitly measured, and then the $k$-means clustering algorithm is
applied to divide the measurements into $k$ measurement groups, and
the wafer coordinates are also clustered according to the $k$
measurement groups.  In the second step, for the subsequently
fabricated wafers, a GP is applied to each cluster
individually. Because the spatial variation is modeled according to
the partitioned magnitude of the measured value, even discontinuous
changes can be reproduced accurately.  In addition, there is a
method~\cite{ITC2016_Butler} dealing with variations between the
sites. In~\cite{ITC2016_Butler}, in order to set outlier limits in a
test, there is a method of eliminating the variation between sites by
normalizing each site~\cite{ITC2016_Butler}. However, it is difficult
to set the normalization constant appropriately in small sampling, and
thus it is difficult to apply the site normalization to the
wafer-level modeling.

However, this method relies heavily on the assumption that the
$k$-means clustering results obtained from the first wafer are
applicable to all subsequent lots. In fact, some site histograms drastically
change in the latest fabrication lot, as shown in
Fig.~\ref{fig:hist07}, which is the histograms in the sixth lot.
For example, the highlighted black histogram should belong to a
different cluster than that shown in Fig.~\ref{fig:hist03} to achieve
accurate prediction. While the possibility of recalibrating $k$-means
clustering is described briefly, no specific solution has been
provided in~\cite{DATE2013_Huang}.

Herein, we propose a novel wafer-level spatial variation modeling
method for RF circuits under multi-site testing. Test engineers
usually have the site information of the probing; thus, we exploit it
as a cluster in the proposed method, to predict spatial variation
through the hierarchical GP modeling of each site.  Therefore, the
proposed method requires no clustering algorithm and no measurement
corresponding to the first step.  The use of site information is
straightforward, but it is efficient under multi-site testing. Because
the characteristics measured within one cluster have the same
additional parasitic components of the calibration circuit, only
spatial changes on the wafer are modeled; as a result, using the
proposed method, accurate modeling can be achieved even across
wafers. We also propose an active sampling method based on {\it active
  learning}~\cite{INNS-ENNS2000_Seo} while considering the measurement
of multi-site testing. Through the active sampling method utilizing
the predictive variance of each site, the proposed method achieves
optimal estimation with a small number of measured samples.

The main contributions of this work are summarized as follows:
\begin{itemize}
  \setlength{\itemsep}{-1pt}
\item {\bf{Hierarchical GP modeling using site information:}} Our method
  enables us to accurately model the spatial correlation on the wafer even
  for the measurement of RF circuits with the discontinuous changes for any
  lots by applying the GP separately to the correct clusters obtained
  from the site information.
\item {\bf{Active sampling algorithm under multi-site testing environment:}}
  We propose an efficient sampling algorithm based on the predictive
  variance of the estimation to determine the sample location.
\item {\bf{Comparison with the conventional method using industrial
    production data:}} We experimentally confirm that the assumption
  of the two-step modeling method~\cite{DATE2013_Huang}, where the
  $k$-means clustering result can be applicable for subsequent wafers, does not
  hold in a more miniaturized fabrication process. We also demonstrate
  that the proposed method can reduce the prediction error to an
  average of $1/19.4$ compared to that obtained using the two-step
  modeling method.
\item {\bf{Thorough evaluation of the proposed active location selection
    algorithm:}} The experimental results also show that the proposed
  sampling method successfully reduces the number of touchdowns
  compared to the random sampling method without sacrificing the
  prediction accuracy. To the best of our knowledge, this is the first
  study to successfully demonstrate spatial variation modeling in a
  multi-site testing environment.
\end{itemize}

The remainder of this paper is organized as
follows. Section~\ref{sec:related} briefly explains GP, which plays a
central role in the proposed method. In addition, we review the
existing wafer-level spatial variation modeling based on the two-step
approach~\cite{DATE2013_Huang}, as a previous work. Then, in
Section~\ref{sec:propose}, a hierarchical GP based on site information
and an active sampling method for multi-site testing are proposed. The
experimental results using industrial production test data of RF IC
fabricated by a 28\,nm process technology are presented in
Section~\ref{sec:exp} to quantitatively evaluate the effectiveness of
the proposed method by comparing it with conventional
methods. Finally, we conclude the paper in
Section~\ref{sec:conclusion}.

\section{Preliminaries}\label{sec:related}
\subsection{Gaussian process}
First, we quickly review a GP~\cite{gp_book}, which is an integral
part of the conventional method~\cite{DATE2013_Huang} and our method.
The GP model is used for estimating the function $y = f(x)$ from the
input variable $x$ to the output variable $y$, which is often used for
regression. In the GP model, the function $f$ is assumed to follow a
multidimensional normal distribution and is expressed as $f \sim
\mathcal{N}(\bm{0},{\bm Z})$ using a kernel matrix ${\bm Z}$.  One of
its advantages is its ability to deal with nonlinear estimation
problems. Another important advantage is the use of Bayesian
inference~\cite{prml}. Because the estimated function is obtained as a
distribution of functions, not as a single function, the uncertainty
of the estimation can be expressed as a predictive variance.

\begin{figure}[t!]
  \begin{algorithm}[H]
    \caption{Gaussian process regression}
    \label{alg:gpr}
    \begin{algorithmic}[1]
      \Require Training dataset: $({\bm X}_{\rm train},{\bm y}_{\rm train})$,
      Test dataset: ${\bm X}_{\rm test}$, Kernel function: $f_{\rm kern}$
      \Ensure Mean and variance of predicted values: ${\bm \mu}=(\mu_1, \mu_2,\cdots, \mu_M)$ and ${\bm v}=(v_1, v_2,\cdots, v_M)$
      \For {$n=1$ to $N$}
      \For {$n'=1$ to $N$}
      \State Calculate $(n,n')$-th element of a kernel matrix ${\bm Z}$ as $f_{\rm kern}({\bm x}_n,{\bm x}_n')$
      \EndFor
      \EndFor
      \State Calculate fitting parameters of $f_{\rm kern}$ to fit $({\bm X}_{\rm train},{\bm y}_{\rm train})$
      \For {$m=1$ to $M$}
      \For {$n=1$ to $N$}
      \State Calculate $n$-th element of ${\bm z_*}$ as $f_{\rm kern}({\bm x}_n,{\bm x^*}_m)$
      \EndFor
      \State ${\bm z_{**}} = f_{\rm kern}({\bm x^*}_m,{\bm x^*}_m)$
      \State Append $\mu_m = {\bm z}^T_* {\bm Z}^{-1}{\bm y}_{\rm train}$ to ${\bm \mu}$
      \State Append $v_m = {\bm z_{**}} - {\bm Z}^{-1}{\bm z}_*$ to ${\bm v}$
      \EndFor
    \end{algorithmic}
  \end{algorithm}
\end{figure}

The outline of a GP-based multiple regression is summarized in
Algorithm~\ref{alg:gpr}. We consider $({\bm X}_{\rm train},{\bm
  y}_{\rm train})=\{({\bm x}_1, y_1), ({\bm x}_2, y_2),\cdots, ({\bm
  x}_N, y_N)\}$ and ${\bm X}_{\rm test}=({\bm x^*}_1, {\bm x^*}_2,
\cdots, {\bm x^*}_M)$ as the training and test datasets, respectively,
where $M \gg N$. In addition, a kernel function $f_{\rm kern}$ is
given as an input. Using the predicted model $f$ calculated based on
$({\bm X}_{\rm train}, {\bm y}_{\rm train})$, the algorithm returns
the mean values and variances of the predicted ${\bm y^*}=(y^*_1,
y^*_2, \cdots, y^*_M)$ for ${\bm X}_{\rm test}$, ${\bm \mu}=(\mu_1,
\mu_2, \cdots, \mu_M)$ and ${\bm v}=(v_1, v_2, \cdots, v_M)$.

In lines 1 to 5, the kernel matrix ${\bm Z}$ of the training dataset
is calculated for each element of ${\bm X}_{\rm train}$ using the
kernel function. Subsequently, in lines 7 to 14, the probability
density function of the predicted $y^*_m$ corresponding to ${\bm
  x^*}_m$ is derived by modeling a multidimensional normal
distribution as follows:
\begin{eqnarray}
  \label{eq:predict}
  & &p(y^*_m|{\bm x^*}_m, {\bm X}_{\rm train}, {\bm y}_{\rm train}) \\ \nonumber
  &=& \mathcal{N}({\bm z}^{\mathrm T}_* {\bm Z}^{-1}{\bm y}_{\rm train}, {\bm z_{**}} -{\bm z}^{\mathrm T}_* {\bm Z}^{-1}{\bm z}_*), 
\end{eqnarray}
where ${\bm z}_*$ and ${\bm z}_{**}$ are the covariances between the
training and test datasets and between the test datasets,
respectively.  As can be seen in Eq.~(\ref{eq:predict}), the mean
value and variance of $y^*_m$ can be analytically derived. The
expected values are used in the prediction, but the variances can also
be used to confirm the uncertainty of the prediction.

There are several kernel functions, such as the linear, squared
exponential, and Mat\'{e}rn kernels. For example,
the radial basis function (RBF) kernel is
as follows~\cite{kernelcookbook,JMLR2001_Genton}:
\begin{eqnarray}
  \label{eq:kernel}
  f_{\rm kern}({\bm x},{\bm x'})=\theta_1 \exp \left( -\frac{({\bm x}-{\bm x'})^2}{\theta_2} \right),
\end{eqnarray}
where $\theta_1$ and $\theta_2$ are the fitting parameters calculated
using an iterative optimization routine, as shown in line 6.
As ${\bm Z}$ is a variance-covariance matrix,
when ${\bm x}$ and ${\bm x'}$ are close, $f_{\rm kern}({\bm
  x},{\bm x'})$ becomes large and, as a result, $f({\bm x})$ and
$f({\bm x'})$ are also close.

The predictive mean $\bm{\mu}$ is used in wafer-level
characteristics modeling.  It can be expected that GP regression can
be incorporated into the wafer-level spatial variation modeling in IC
characteristics with high affinity, as the characteristics of adjacent
dies on the wafer are similar because of the systematic components of
process variation~\cite{TSM2004_Ohkawa,TED2008_Saxena}.

\subsection{Related work}
Owing to the intensive research on wafer-level spatial variation
correlation modeling, the prediction accuracy of the spatial
measurement variation has been improved, thereby, enabling the
successful reduction of measurement costs in production
tests~\cite{TSM2010_Reda,ICCAD2009_Li,DAC2010_Zhang,TCAD2011_Zhang,DATE2014_Zhang,ICCAD2012_Kupp,ITC2014_Ahmadi,DATE2013_Huang}.
Among others, in~\cite{DATE2013_Huang}, a two-step modeling approach
has been proposed to handle the discontinuous effect induced by
multi-site testing and reticle shot, etc., in wafer-level modeling.

The objective of the first step is to partition the wafer into
$k$ groups, which reflect the $k$ levels of wafer measurement induced
by discontinuous effects. For this purpose, the $k$-means algorithm is
exploited as follows:
\begin{eqnarray}
  {\bm y}=\{{\bm y}^{(1)}, {\bm y}^{(2)}, \cdots, {\bm y}^{(k)}\},
  \label{eq:1st_y}
\end{eqnarray}
where ${\bm y}$ represents the vector of the measured characteristics
of all the dies on the wafer. Consequently, ${\bm X}$ corresponding
to ${\bm y}$ is partitioned as:
\begin{eqnarray}
  {\bm X}=\{{\bm X}^{(1)}, {\bm X}^{(2)}, \cdots, {\bm X}^{(k)}\}.
  \label{eq:1st_x}
\end{eqnarray}
Note that Eq.~(\ref{eq:1st_x}) indicates that the coordinates on the
wafer are divided according to the measured characteristics.  Once the
$k$ clusters are identified, in the second step for subsequent wafers,
the GP is applied to each cluster individually based on
Algorithm~\ref{alg:gpr}. Because the changes in each ${\bm y}^{(k)}$
can be expected to be smooth, the GP regression will work
successfully, and thus the two-step approach can handle discontinuous
changes.

Note that determining the optimal $k$ is not trivial.  Although
several methods, such as silhouette value~\cite{JCAM1987_Rousseeuw}
and the elbow method~\cite{NeuroImage1999_Goutte}, are well known to
determine optimal $k$, in~\cite{DATE2013_Huang}, $k$ is determined
based on the following equation:
\begin{eqnarray}
  k = \argmax_g\, CH(g),
  \label{eq:ch}
\end{eqnarray}
where $CH(g)$ is the Calinski and Harabasz index when $g$ clusters are
considered~\cite{CS1974_Calinski}.

However, the two-step modeling is not always applicable because it
keeps using the $k$ clusters for subsequent wafers, assuming that the
content of the clusters will not change for other wafers/lots. Because
the experiment in~\cite{DATE2013_Huang} uses the industrial data
fabricated using a relatively mature process technology, the
assumption might hold true; in contrast, for our production data on
immature process technology, the process is inapplicable, as shown in
Fig.~\ref{fig:hist}.

\begin{figure}[t!]
  \centering
  \includegraphics[width=.83\linewidth]{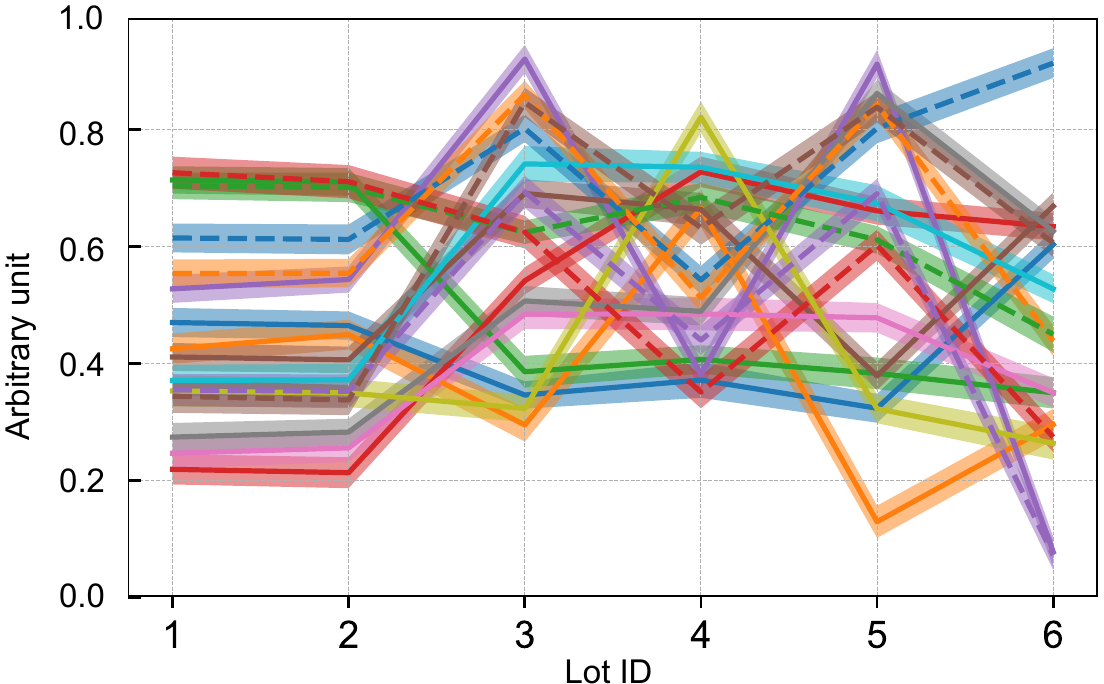}
  \caption{Measured characteristics of 16 sites from the first lot to
    the sixth lot. The solid lines and shaded regions represent the
    means and the three standard deviations of the variations,
    respectively. The vertical axis denotes arbitrary units.}
  \label{fig:lot-lot}
\end{figure}

Another observation in Fig.~\ref{fig:hist} is shown in
Fig~\ref{fig:lot-lot}.  In this figure, the measured characteristics
for each site are shown as functions of the lot ID from the first lot
to the sixth lot. Here, the first wafer is used for each lot. The
lines and shaded regions represent the mean and three standard
deviations of each site, respectively. It can be seen that the
distributions within the site are comparatively maintained up to the
first two lots; however, they fluctuate greatly from the third
lot. This suggests that the two-step modeling may work well up to the
first two lots, whereas the clusters need to be recalibrated for the
third and sixth lots, thus resulting in additional measurement
costs. In addition, the early stage lots generally have low production
yields, making it difficult to apply the two-step modeling method.

\section{Wafer-level variation modeling for RF IC under multi-site testing}\label{sec:propose}
We propose a novel spatial variation model based on the site
information provided by test engineers, which can give us the correct
cluster without applying clustering algorithms. The GP-based
prediction is hierarchically performed for each site cluster.
Site-to-site variations are caused by parasitic components in the
calibration circuit during multi-site testing.  Ideally, they should
be eliminated during measurement, which is impractical due to the
design and manufacturing costs of the probe card.  They also can be
solved by considering them one at a time, but the benefits of
multi-site testing cannot be obtained. The proposed method applies
hierarchical GP modeling by clustering using site information and
achieves highly accurate modeling while considering the actual
measurement environment. As observed in Fig.~\ref{fig:lot-lot}, the
measurements at the same site have a small deviation. The site-based
hierarchical clustering can always be expected to be a good model
without recalibration.

In addition, we propose an active sampling algorithm to achieve a
small sampling ratio based on variance computed by GP-based regression
in a multi-site testing environment. In contrast to all the existing
studies that assume sampling one at a time, the proposed algorithm can
effectively reduce the measurement cost.

\subsection{Modeling based on site-based hierarchical GP}

\begin{figure}[t!]
  \begin{algorithm}[H]
    \caption{Site-based hierarchical spatial variation modeling}
    \label{alg:prop}
    \begin{algorithmic}[1]
      \Require ${\bm \mu}$ and ${\bm v}$
      Training dataset: $({\bm X}_{\rm train},{\bm y}_{\rm train})$ measured under multi-site testing,
      Test dataset: ${\bm X}_{\rm test}$, Kernel function: $f_{\rm kern}$, site information
      \Ensure Mean and variance of predicted values: ${\bm \mu}=(\mu_1, \mu_2,\cdots, \mu_M)$ and ${\bm v}=(v_1, v_2,\cdots, v_M)$
      \State Cluster $({\bm X}_{\rm train},{\bm y}_{\rm train})$ and ${\bm X}_{\rm test}$ into $S$ groups according to the site information
      \For {$s=1$ to $S$}
      \State ${\bm \mu}^{(s)}$, ${\bm v}^{(s)}$ = gpr$\left(({\bm X}^{(s)}_{\rm train}, {\bm y}^{(s)}_{\rm train}), {\bm X}^{(s)}_{\rm test}, f_{\rm kern}\right)$
      \EndFor
      \State Concatenate all ${\bm \mu}^{(s)}$ and ${\bm v}^{(s)}$ into ${\bm \mu}$ and ${\bm v}$
    \end{algorithmic}
  \end{algorithm}
\end{figure}

Algorithm~\ref{alg:prop} presents the proposed spatial correlation
modeling through a site-based hierarchical GP in detail.  We assume
that the measurement is conducted by multi-site testing. The
distinction from the conventional method is that the clustering is
performed according to the site information in a single touchdown as
listed in line 1, i.e., the conventional method needs to measure the
characteristics of an entire wafer, whereas the proposed method
requires no measurement for clustering. In Algorithm~\ref{alg:prop},
$S$ represents the number of the sites in a single touchdown, and the
training and test datasets are grouped into $S$ groups as follows:
\begin{eqnarray}
  ({\bm X}_{\rm train}, {\bm y}_{\rm train})=\{({\bm X}^{(1)}_{\rm train}, {\bm y}^{(1)}_{\rm train}), ({\bm X}^{(2)}_{\rm train}, {\bm y}^{(2)}_{\rm train}), \nonumber \\
  \cdots, ({\bm X}^{(S)}_{\rm train}, {\bm y}^{(S)}_{\rm train})\}
  \label{eq:site_train}
\end{eqnarray}
and
\begin{eqnarray}
  {\bm X}_{\rm test} = \{{\bm X}^{(1)}_{\rm test}, {\bm X}^{(2)}_{\rm test}, \cdots, {\bm X}^{(S)}_{\rm test}\},
  \label{eq:site_test}
\end{eqnarray}
respectively.  The GP-based regression is performed individually by
modeling each site hierarchically as listed in lines 2 to 4 based on
the {\it gpr} function listed in Algorithm~\ref{alg:gpr}, where the
mean and variance of the prediction for the test dataset are returned.
Finally, the prediction result for the entire wafer is obtained
through the concatenation of each prediction result.

\begin{figure}[!t]
  \centering \subfigure[Eight sampled touchdown locations in
    multi-site testing]{
    \includegraphics[width=.6\linewidth]{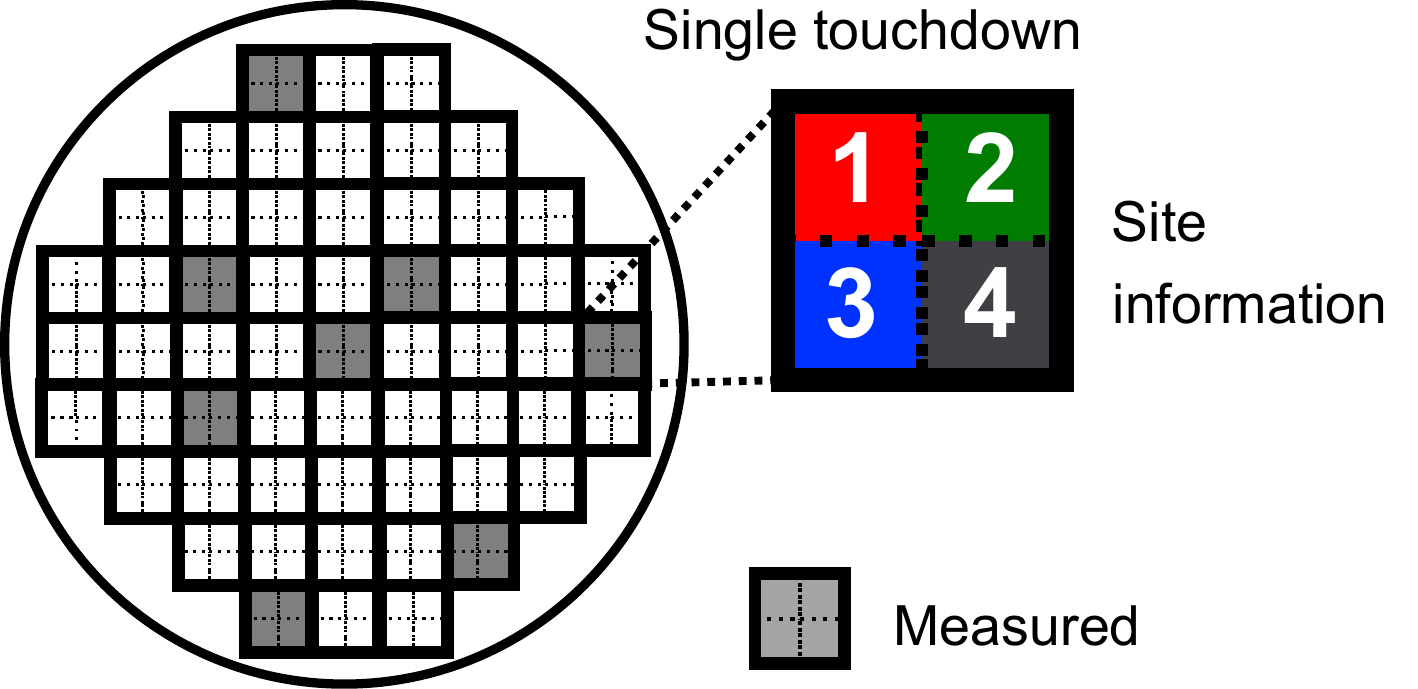}
    \label{fig:1st}
  }
  \subfigure[Modeling and prediction using hierarchical GP regression]{
    \includegraphics[width=.65\linewidth]{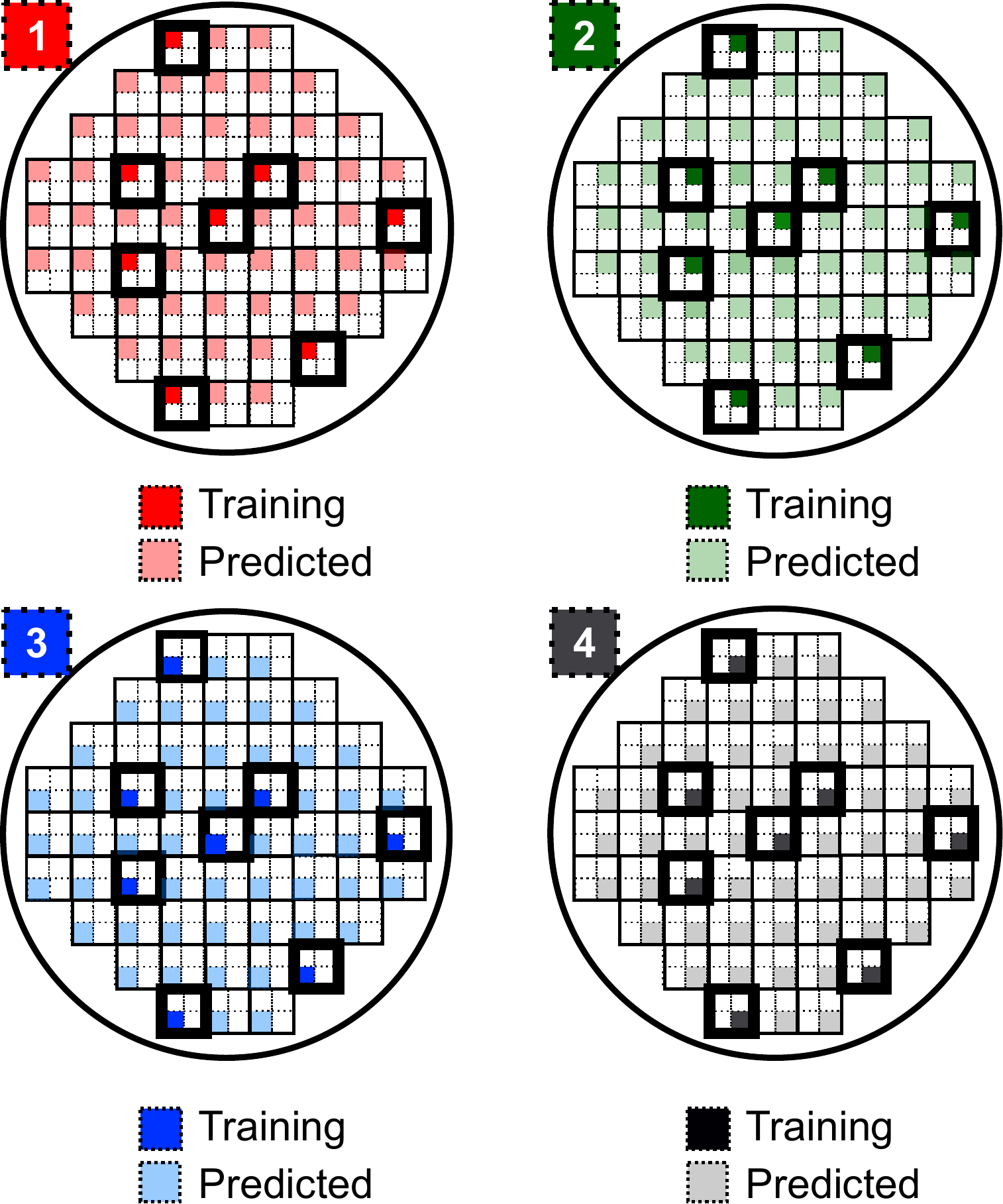}
    \label{fig:2nd}
  }\\
  \subfigure[All-site concatenation]{
    \includegraphics[width=.32\linewidth]{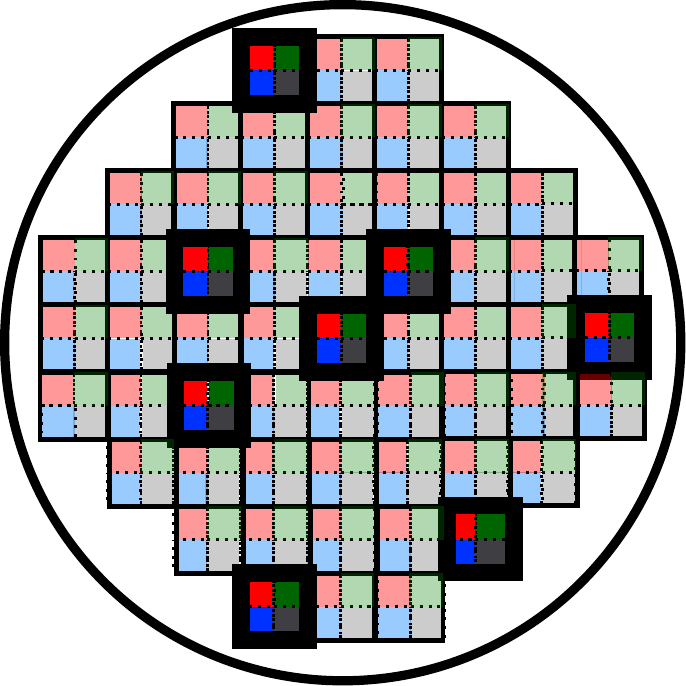}
    \label{fig:3rd}
  }
  \caption{Example of the site-based hierarchical GP regression, where
    a single touchdown has four sites.}
  \label{fig:prop}
\end{figure}

Figure~\ref{fig:prop} depicts an example of the modeling using the
proposed method when $S=4$ (sites 1 to 4).  First, the eight positions
are selected and measured as the training as shown in
Fig.~\ref{fig:1st}. In total, 32 dies ($=8 \times 4$) are measured
using the eight touchdowns. Then, according to the site, the GP-based
modeling and prediction are individually applied, as in
Fig.~\ref{fig:2nd}. In other words, for site~1, the measured value
belonging to site~1 is used as training data to build a GP model, and
the measured value of the unmeasured die belonging to site~1 is
predicted. The measurements and predictions for the other sites are
also similar. The entire prediction result is obtained through the
concatenation as shown in Fig.~\ref{fig:3rd}.

\subsection{Active sampling under the multi-site testing}\label{sec:sampling}
In the wafer-level spatial modeling, it is desirable to be given a
small input training dataset to maximize the cost reduction of the
measurement.  In~\cite{ITC2014_Ahmadi}, an aggressive sampling method
that preferentially measures the location with the largest predictive
variance calculated by the GP regression is proposed. In addition,
in~\cite{TCAD2011_Zhang}, a Latin hypercube sampling
approach~\cite{JASA1991_Tang} is employed to evenly choose random
sample points over the entire wafer.  However, these methods are
straightforward, and most importantly, they do not consider
multi-site testing environment.

A good model should be one with a small error between the model and
the actual measurement.  The mean squared error (MSE) against the test
dataset can be expressed as:
\begin{eqnarray}
  E_{\rm MSE} = ||{\bm v}|| + ||{\bm \mu} - {\bm y}_{\rm true}||^2,
  \label{eq:rmse}
\end{eqnarray}
where $||\cdot||$ is the Euclidean norm, and ${\bm y}_{\rm true}$ is
the correct value at the location of ${\bm X}_{\rm test}$ and unknown,
i.e., unmeasured value. Assuming that the model is correct, the
contribution of the second term in Eq.~(\ref{eq:rmse}) to $E_{\rm
  MSE}$ is small compared to the variance contribution, that is, the
first term. Thus, to minimize $E_{\rm MSE}$, we have to select ${\bm
  X}$ such that the overall variance of the estimator is
minimized~\cite{INNS-ENNS2000_Seo}.

\begin{figure}[t!]
  \begin{algorithm}[H]
    \caption{Active location selection with site-based hierarchical Gaussian process regression}
    \label{alg:active}
    \begin{algorithmic}[1]
      \State ${\bm \mu}$, ${\bm v}$ = hgpr$\left(({\bm X}_{\rm train}, {\bm y}_{\rm train}), {\bm X}_{\rm test}, f_{\rm kern}\right)$
      \For {$p=1$ to $P$}
      \State ${\bm \mu}_p$, ${\bm v}_p$ = hgpr(${\bm X}_{\rm train}+{\bm X}^{(p)}_{\rm add}, {\bm X}_{\rm test}, f_{\rm kern}$)
      \State Calculate the Euclidean distance between ${\bm v}$ and ${\bm v}_p$ as ${\Delta}^{(p)}_{{\rm var}}$
      \EndFor
      \State Select ${\bm X}_p$ with the largest ${\Delta}^{(p)}_{{\rm
    var}}$ as the next touchdown location
    \end{algorithmic}
  \end{algorithm}
\end{figure}

Based on the aforementioned discussion, we propose an active sampling
method as outlined in Algorithm~\ref{alg:active}, which is
incorporated into the site-based hierarchical spatial modeling ({\it
  hgpr}) shown in Algorithm~\ref{alg:prop}. The proposed sampling
method focuses on the Euclidean distance between before and after
measurements.
The proposed method proceeds as follows. The numbers on the left
indicate the corresponding line numbers in Algorithm~\ref{alg:active}.
\begin{itemize}
\item[1)] Calculate ${\bm \mu}$ and ${\bm v}$ through the
  hierarchical GP regression using $({\bm X}_{\rm train}, {\bm y}_{\rm
    train})$ and ${\bm X}_{\rm test}$ as in
  Algorithm~\ref{alg:prop}. ${\bm X}_{\rm train}$ can be obtained through
  multi-site testing.
\item[2)] Repeat steps 3) and 4) for all touchdown
  location candidates. Here, the $p$-th touchdown candidate has ${\bm
    X}^{(p)}_{\rm test}=\{{\bm x}^{*(p)}_{1}, {\bm x}^{*(p)}_{2},
  \cdots, {\bm x}^{*(p)}_{S}\}$ with the $S$ sites.
\item[3)] Add the touchdown candidate ${\bm X}^{(p)}_{\rm add}$ by
  assuming it as measured and perform the hierarchical
  GP regression. Note that because it is not actually measured, we
  assume the mean values are measured as ${\bm X}^{(p)}_{\rm
    add}=\{({\bm x}^{*(p)}_{1}, \mu^{(p)}_{1}), ({\bm x}^{*(p)}_{2},
  \mu^{(p)}_{2}), \cdots, ({\bm x}^{*(p)}_{S}, \mu^{(p)}_{S})\}$,
  where $\mu^{(p)}_{S}$ is the predicted mean corresponding to ${\bm
    x}^{*(p)}_{S}$, and is one of the elements of ${\bm \mu}$ given by
  hgpr in step 1). In this step, ${\bm \mu}_p$ and ${\bm v}_p$ are
  obtained as in step 1).
\item[4,5)] Calculate the Euclidean distance of ${\bm v}$ and ${\bm
  v}_p$ as ${\Delta}^{(p)}_{{\rm var}}$.  Note that steps 2) to 5) are
  iterated for all the touchdown candidates.
\item[6)] Select ${\bm X}_p$ with the largest ${\Delta}^{(p)}_{{\rm
    var}}$ as the next measurement location.
\end{itemize}
The mentioned procedure is iterated until an exit condition is
satisfied, for example, a sufficient number of iterations are
obtained. Because the reduction of the whole deviation for the test
dataset is compared in step 6), a more accurate modeling can be
expected with a smaller number of measurements compared to simply
checking the location of the highest variance according
to~\cite{INNS-ENNS2000_Seo}.

\section{Numerical experiments}\label{sec:exp}
\begin{figure}[t!]
  \centering
  \includegraphics[width=.57\linewidth]{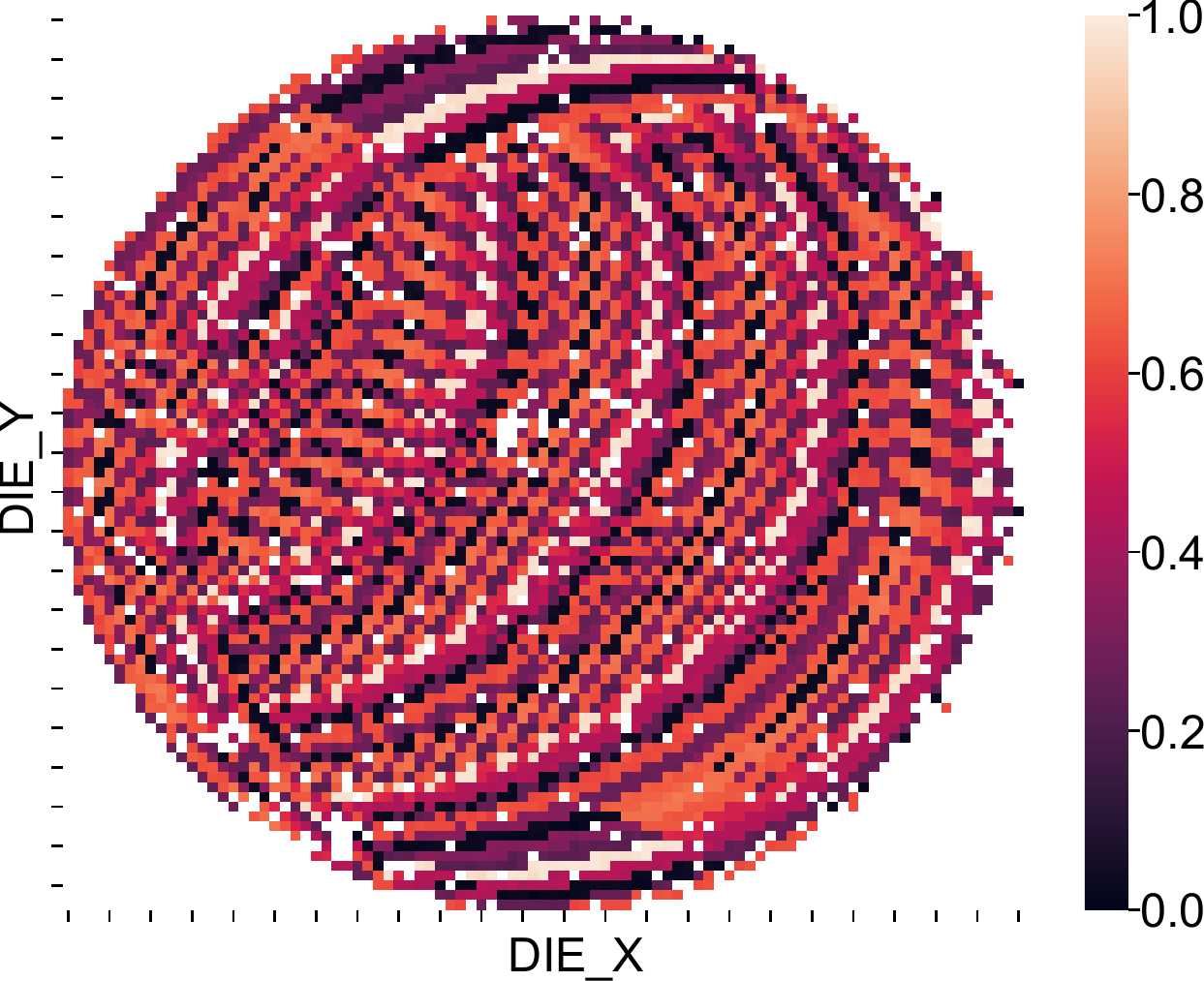}
  \caption{Heat map of fully measured characterization. The measured
    values are normalized.}
  \label{fig:full}
\end{figure}

\begin{figure}[t!]
  \centering
  \includegraphics[width=0.5\linewidth]{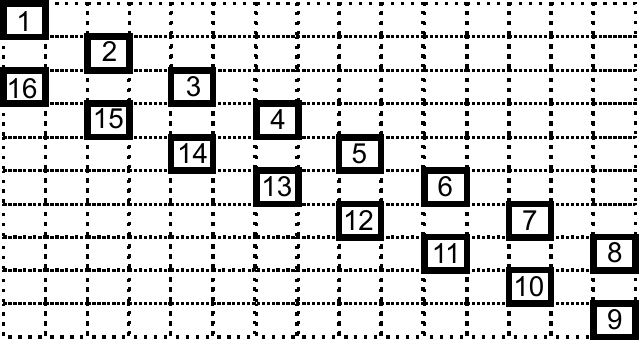}
  \caption{Single touchdown with 16 sites in our multi-site testing.}
  \label{fig:td}
\end{figure}

\subsection{Setup}
To demonstrate the effectiveness of the proposed method, we conducted
experiments using an industrial production test dataset of a 28\,nm
analog/RF device. Our dataset contains six lots. The first wafer of
each lot was used for the evaluation; thus, we used six wafers with
different lots. A single wafer has approximately 6,000 DUTs. In this
experiment, we used a measured character for an item of the dynamic
current test, in which site-to-site variability due to the multi-site
test is noticeably observed, as shown in Figs.~\ref{fig:hist}
and~\ref{fig:lot-lot}. A heat map of the full measurement results for
the first wafer of the sixth lot is shown in Fig.~\ref{fig:full}. For
the ease of experimentation, faulty dies were removed from the
dataset. The number of sites in a single touchdown is 16, i.e.,
$S=16$. The form of a single touchdown is illustrated in
Fig.~\ref{fig:td}. This is different from the rectangular touchdown
illustrated in Fig.~\ref{fig:prop}, which prevents interference on the
probe of the impedance-matching circuits. As a result, a special
pattern is observed during the multi-site test as shown in
Fig.~\ref{fig:full}. To fully measure all DUTs on a single wafer,
approximately 600 touchdowns are required.

All experiments were implemented in the Python language.  The RBF
kernel was used as the kernel function $f_{\rm kern}$ for the GP-based
regression.  The experiments were conducted on a Linux PC with an
Intel Xeon Platinum 8160 2.10\,GHz central processing unit using a
single thread.

To quantitatively evaluate the modeling accuracy, we define the
error ${\bm \delta}$ between the correct ${\bm y}_{\rm true}$ and the
predicted mean ${\bm \mu}$ normalized by the maximum and minimum
values of ${\bm y}_{\rm true}$ as follows:
\begin{eqnarray}
  {\bm \delta} = \frac{{\bm \mu}-{\bm y}_{\rm true}}
  {d_{\rm spec}},
  \label{eq:delta}
\end{eqnarray}
where $d_{\rm spec}$ is the range between the minimum and maximum
values of the fully measured characteristics in Fig.~\ref{fig:full}.

\subsection{Experimental results on site-based hierarchical spatial modeling}
We first evaluated the site-based hierarchical spatial modeling
presented in Algorithm~\ref{alg:prop}.  For comparison, naive GP
regression-based approach (hereafter called {\it naive
  GP})~\cite{ICCAD2012_Kupp} and the two-step approach (hereafter
called {\it 2-step GP})~\cite{DATE2013_Huang} are also applied. Here,
it should be noted that we do not consider the touchdown, that is,
one-by-one measurement is conducted. The experimental result based on
the touchdown is described later in Section~\ref{sec:active_res}.

\begin{figure}[t]
  \centering \subfigure[Naive GP]{
    \includegraphics[width=.57\linewidth]{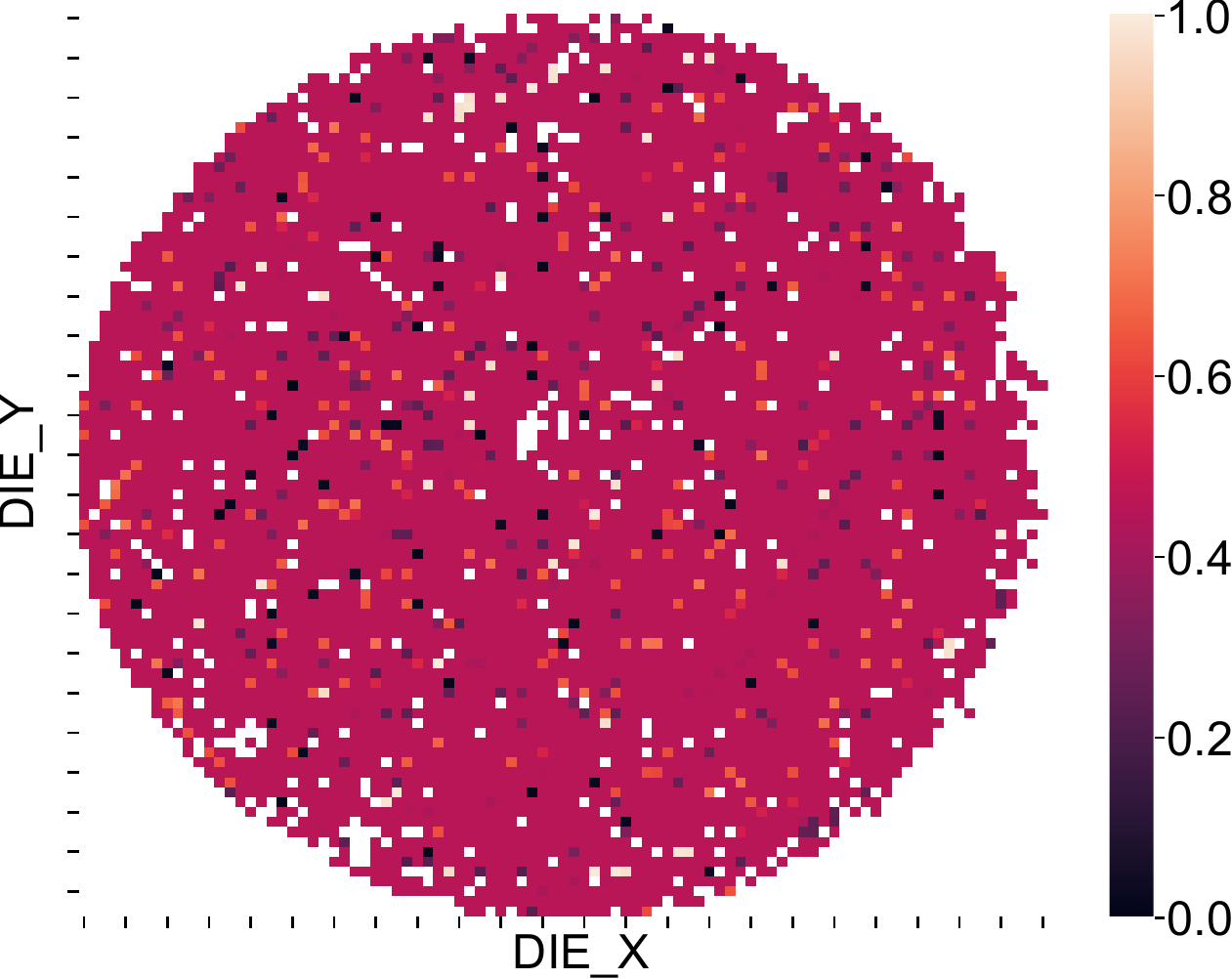}
    \label{fig:ngp}
  }
  \hspace{-4mm}
  \subfigure[2-step GP]{
    \includegraphics[width=.57\linewidth]{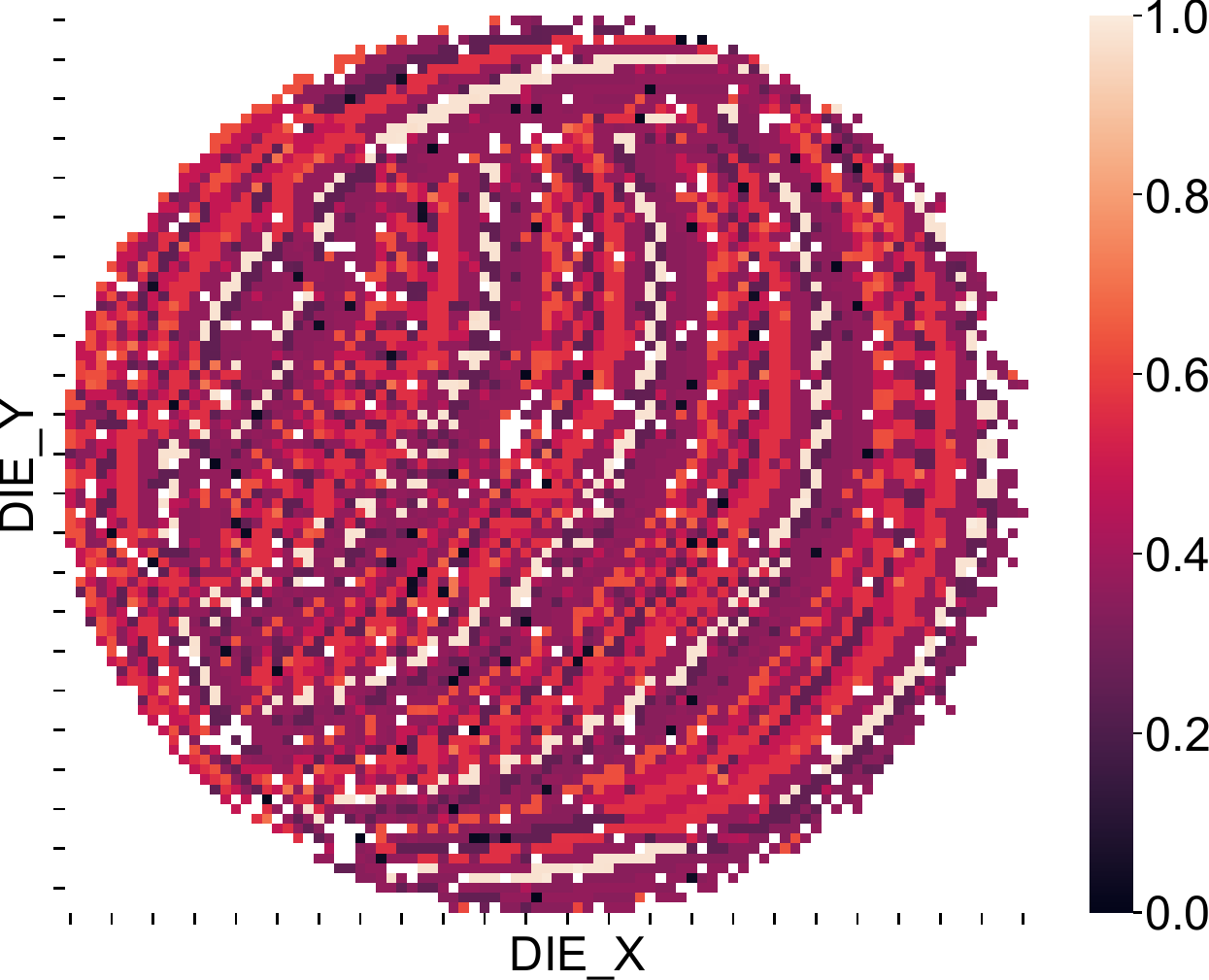}
    \label{fig:2gp}
  }
  \subfigure[Ours]{
    \includegraphics[width=.57\linewidth]{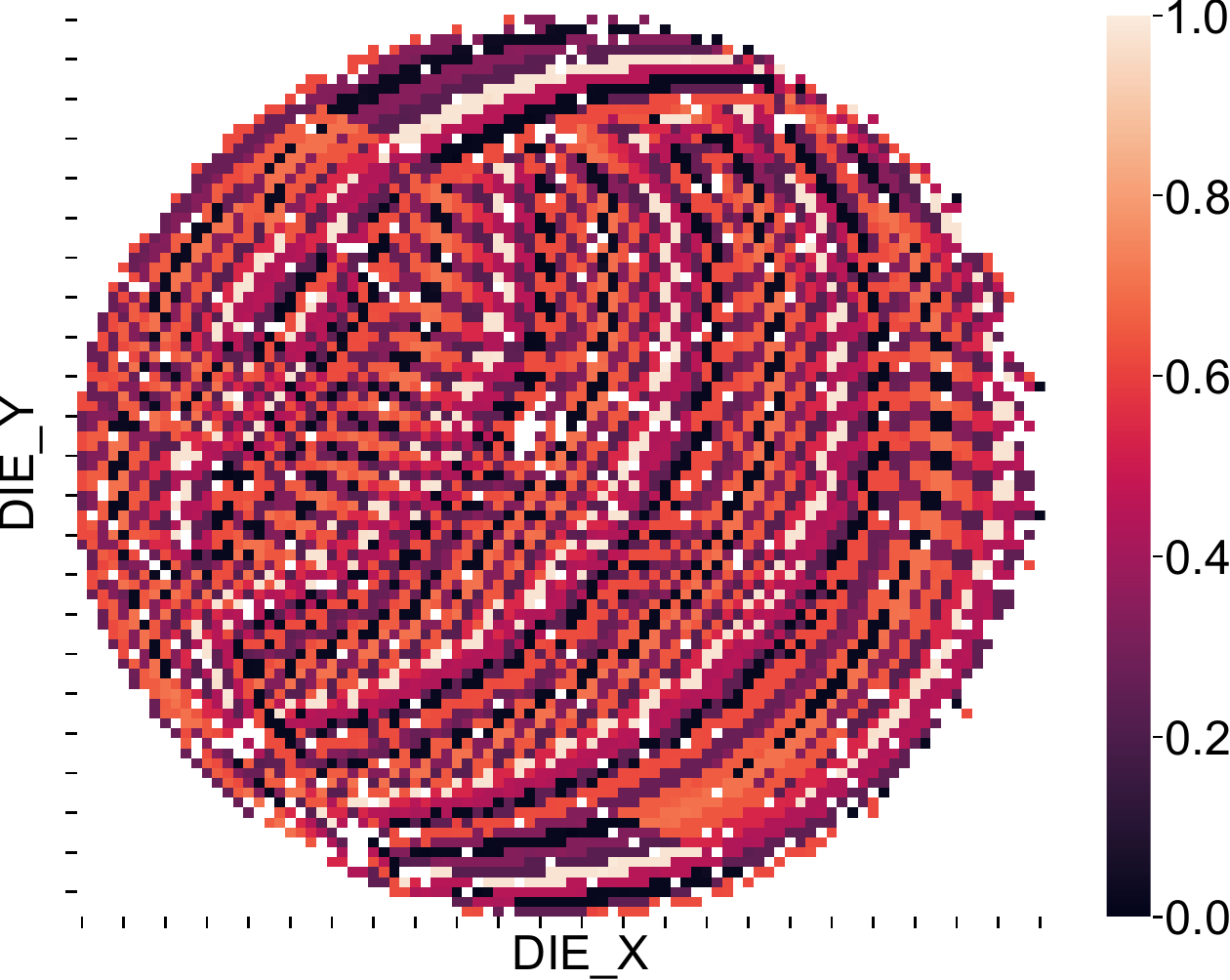}
    \label{fig:ours}
  }
  \caption{Heat maps of the predicted characteristics by naive GP,
    2-step GP, and the proposed method at the 10\% spatial sampling
    rate. It can be visibly confirmed that the prediction results are
    closer to the actual measurements in the order of naive GP, 2-step
    GP, and the proposed method. The measured values are normalized.}
  \label{fig:phase1}
\end{figure}

For the 2-step GP method, the first wafer of the first lot is used to
obtain $k$ clusters through $k$-means clustering. For the
subsequent wafers, $k$ clusters were used to predict the device
characteristics. In the experiment, the optimal $k$ was determined
using the silhouette value and elbow
method~\cite{JCAM1987_Rousseeuw,NeuroImage1999_Goutte} instead of
Eq.~(\ref{eq:ch}), resulting in seven clusters.

\begin{figure}[t!]
  \centering
  \includegraphics[width=.65\linewidth]{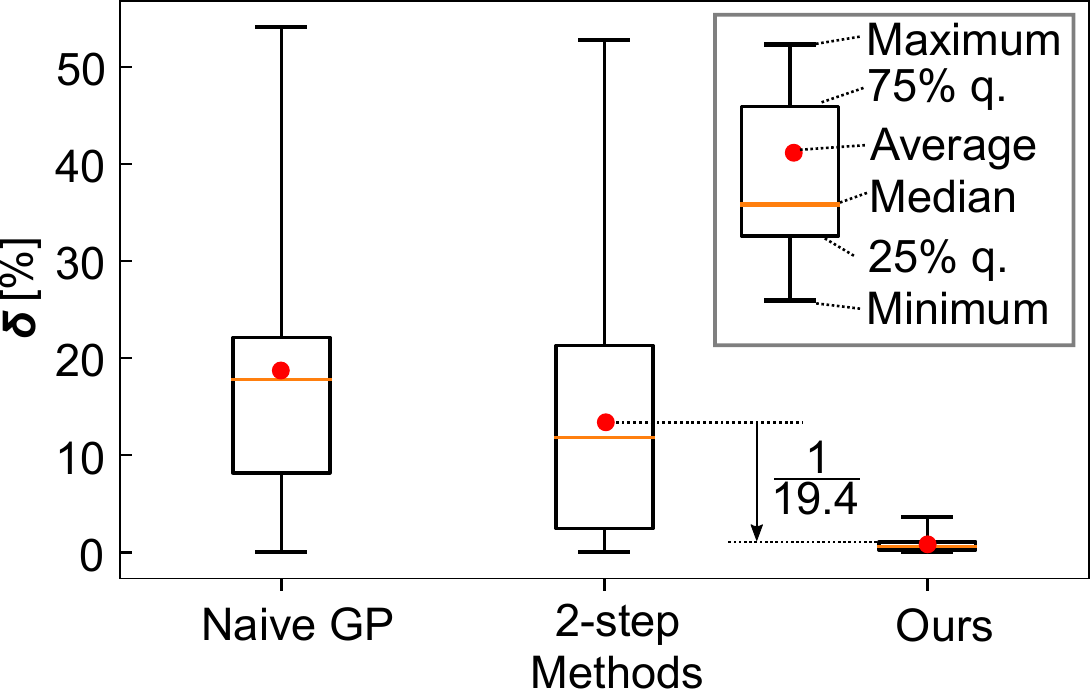}
  \caption{Box-plots of ${\bm \delta}$ for each method, where the
    maximum, 75\% quantile, median, average, 25\% quantile, and
    minimum values are shown.}
  \label{fig:p1_boxplot}
\end{figure}

In Fig.~\ref{fig:phase1}, the prediction results for the wafer of the
sixth lot using each method are shown. They were predicted using
randomly sampled values at a 10\% spatial sampling rate. Clearly,
the naive GP method fails to capture the site-to-site variation as shown
in Fig.~\ref{fig:ngp}. In contrast, the specific pattern shown in
Fig.~\ref{fig:full} caused by the site-to-site variation can be
confirmed in the 2-step GP method and our method as shown in
Figs.~\ref{fig:2gp} and~\ref{fig:ours}.
Figure~\ref{fig:p1_boxplot} shows the box-plots of ${\bm \delta}$
using Eq.~(\ref{eq:delta}) for each method.  In the figure, the top
and bottom of the line represent the maximum and minimum values,
respectively. The average is shown as a dot. The top and bottom of the
box are 75\% quantile and 25\% quantile, respectively. The line
represents the median. The average errors of ${\bm \delta}$ of the
naive GP, 2-step GP method, and our method are 18.59\%, 13.43\%, and
0.69\%, respectively.  The proposed method can also drastically reduce
the variance in the predictions.  From the results, we can conclude
that the proposed method can reduce the average error by approximately
5.13\% compared to the 2-step GP method, i.e., $19.46$ times
$(=13.43/0.69)$ more accurately.

\begin{figure}[t!]
  \centering
  \includegraphics[width=.7\linewidth]{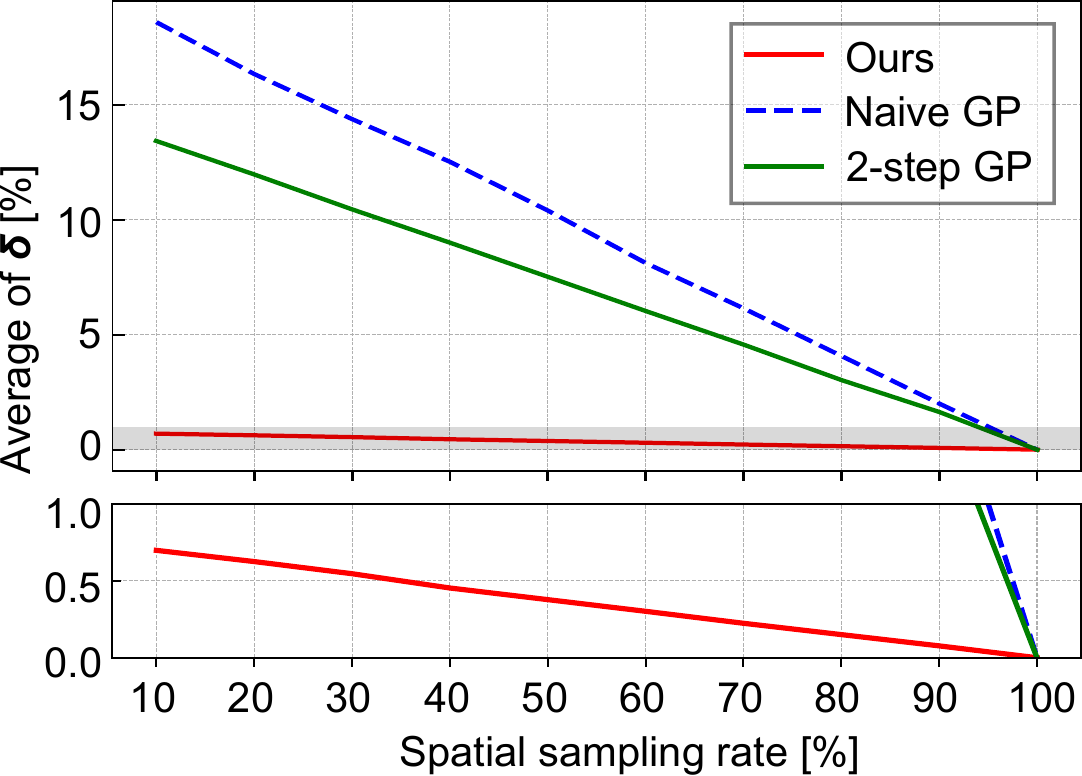}
  \caption{Averages of ${\bm \delta}$ obtained by naive GP, 2-step GP,
    and the proposed method at various sampling rates. The gray part is enlarged at the bottom of this figure.}
  \label{fig:p1_lot7}
\end{figure}

Figure~\ref{fig:p1_lot7} shows the averages of ${\bm \delta}$ as a
function of the spatial sampling rate using the three methods for the
wafer of the sixth lot.  Note that the prediction methods are not
applied when the spatial sampling rate is 100\% and the sampling rate
is incrementally increased, that is, the measured locations at the
10\% sampling rate are always contained at subsequent rates. As the
spatial sampling rate increases, the averages of all the methods
decrease monotonically. We also find that the average errors of the
proposed method always achieves better prediction results for all the
sampling rates.

\begin{figure}[t!]
  \centering
  \includegraphics[width=.65\linewidth]{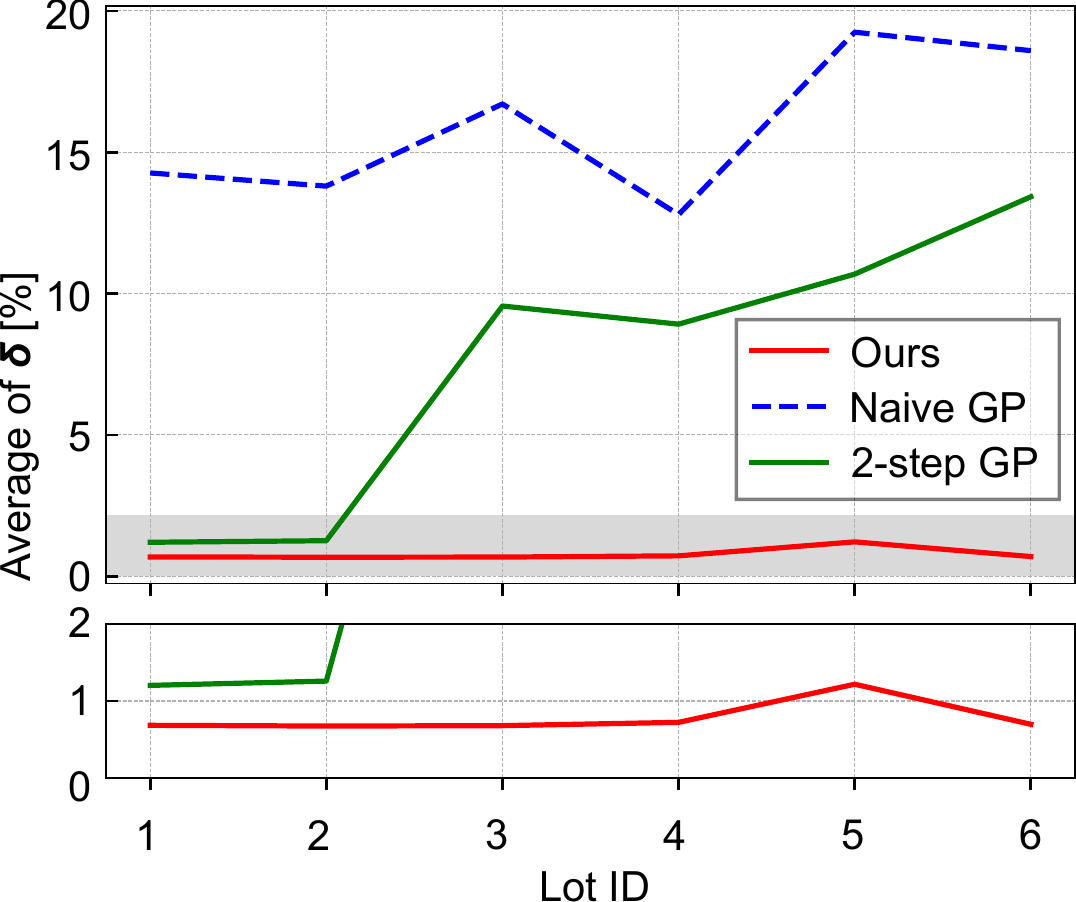}
  \caption{Changes of the averages of ${\bm \delta}$ for each lot.
    The gray part is enlarged at the bottom of this figure.}
  \label{fig:p1_lot1-lot7}
\end{figure}

The averages of ${\bm \delta}$ for the wafer at the 10\% sampling rate
as a function of the lot ID are shown in
Fig.~\ref{fig:p1_lot1-lot7}. From the figure, it can be observed that
the proposed method achieves the best estimation results for all the
lots among the three methods. The prediction performance of the 2-step
GP degrades as the production lot progresses, while the proposed
method maintains a low prediction error below 2\% regardless of the
lot.  This result implies that the $k$-means clustering result
obtained in the first lot is inappropriate for subsequent lots.

\begin{figure}[t!]
  \centering
  \includegraphics[width=.75\linewidth]{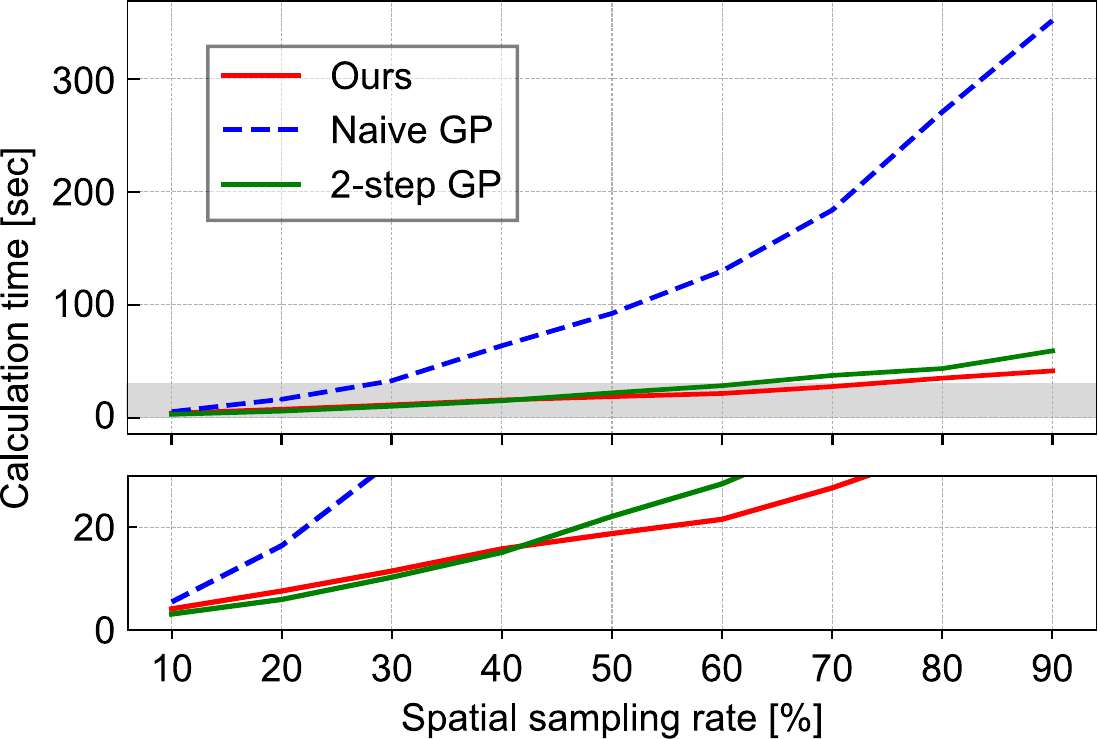}
  \caption{Calculation time. The gray part is enlarged at the bottom of this figure.}
  \label{fig:time}
\end{figure}

We evaluated the calculation time of the prediction for each method.
Figure~\ref{fig:time} summarizes the calculation time of each method
for the sixth lot at various sampling rates.  We can see that the
proposed method and 2-step method can significantly reduce the
calculation time compared to the naive GP method.  This is due to the
side benefits of hierarchical GP modeling approaches.  In general, the
inference time of GP is $\mathcal{O}(N^3)$ scaled because of the
computation of the matrix
inverse~\cite{ACML2010_Park,SAC2019_Nguyen}. In the proposed method,
GP modeling is conducted for each site individually, and thus the
calculation time can be drastically reduced because the training
samples are reduced to $N/S$ in each GP modeling, where $S$ is 16 for
the proposed. The reduction becomes $N/k$ for the 2-step method, where
$k=7$.  Note that this calculation was conducted using a single
thread. Therefore, the calculation time can be further reduced by
implementing parallel processing.

\subsection{Experimental result under multi-site testing}\label{sec:active_res}
In the evaluation of the previous section, the sample dies are
randomly selected one by one, and thus the multi-site test environment
that measurements are conducted per site unit is not considered.  We
evaluated the sampling method listed in Algorithm~\ref{alg:active}
under the multi-site test environment. Here, it is assumed that the
touchdown shown in Fig.~\ref{fig:td} is performed in a single
measurement. In all the existing researches of the wafer-level
variation modeling, sampling is assumed one DUT at a time, and thus
this work is the first to consider a multi-site testing environment
for wafer-level variation modeling. In this experiment, the random
sampling was used for comparison. First, we measured one randomly
sampled touchdown and then selected the next one by the proposed
method and random sampling.

\begin{figure}[t!]
  \centering
  \includegraphics[width=.7\linewidth]{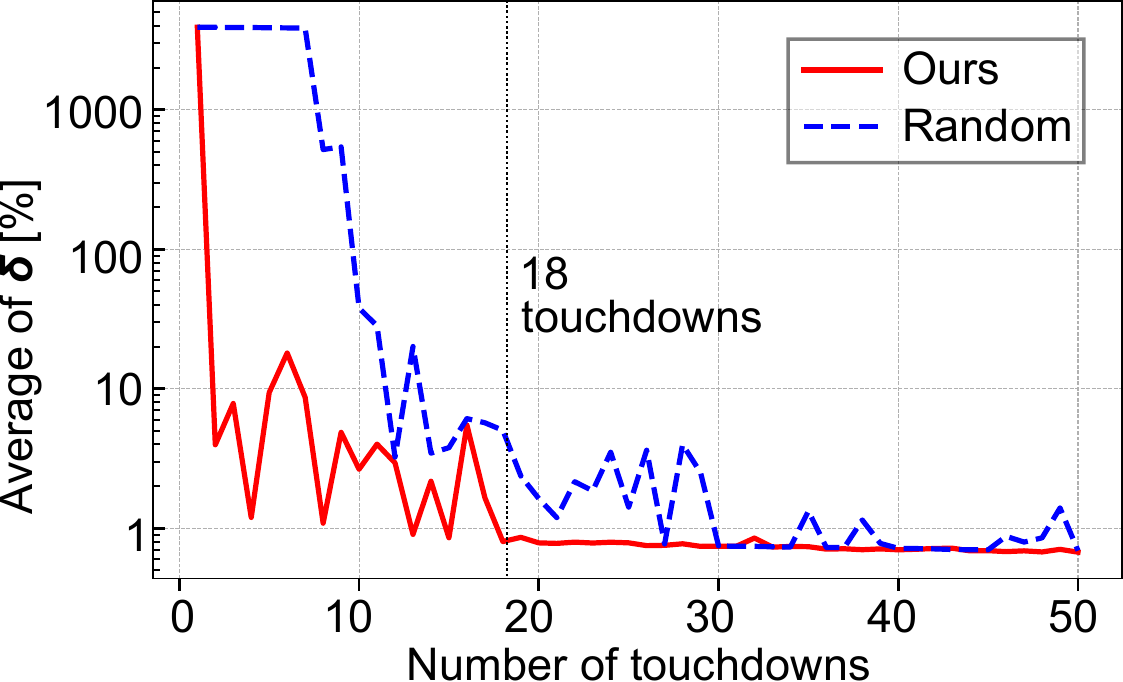}
  \caption{Averages of ${\bm \delta}$ as a function of the number of
    the touchdowns. The vertical axis is shown in log scale. The
    proposed method converges more quickly.}
  \label{fig:p2_average}
\end{figure}

Figure~\ref{fig:p2_average} shows the average of ${\bm \delta}$ as a
function of the number of the touchdowns that incrementally
increased. The first wafer in the sixth lot was used.  Though, an
error of over 3000\% is observed for both the methods at the first
touchdown, the error is decreasing as increasing the number of the
touchdowns for both methods.  However, the random sampling method
converges slowly, whereas the proposed method converges further
quickly.  More specifically, the proposed method has an average error
of 0.80\% or less with the 18 touchdowns, in contrast, the random
sampling still has an average of 5.03\%. The 18 touchdowns correspond
to approximately 3\% of the number of the touchdowns for full
measurement. In addition, the random sampling does not converge even
at the 50 touchdowns.  It can be seen from the figure that the
proposed sampling method can successfully reduce the number of the
necessary touchdowns while achieving better prediction accuracy.

\begin{figure}[t!]
  \centering
  \includegraphics[width=.65\linewidth]{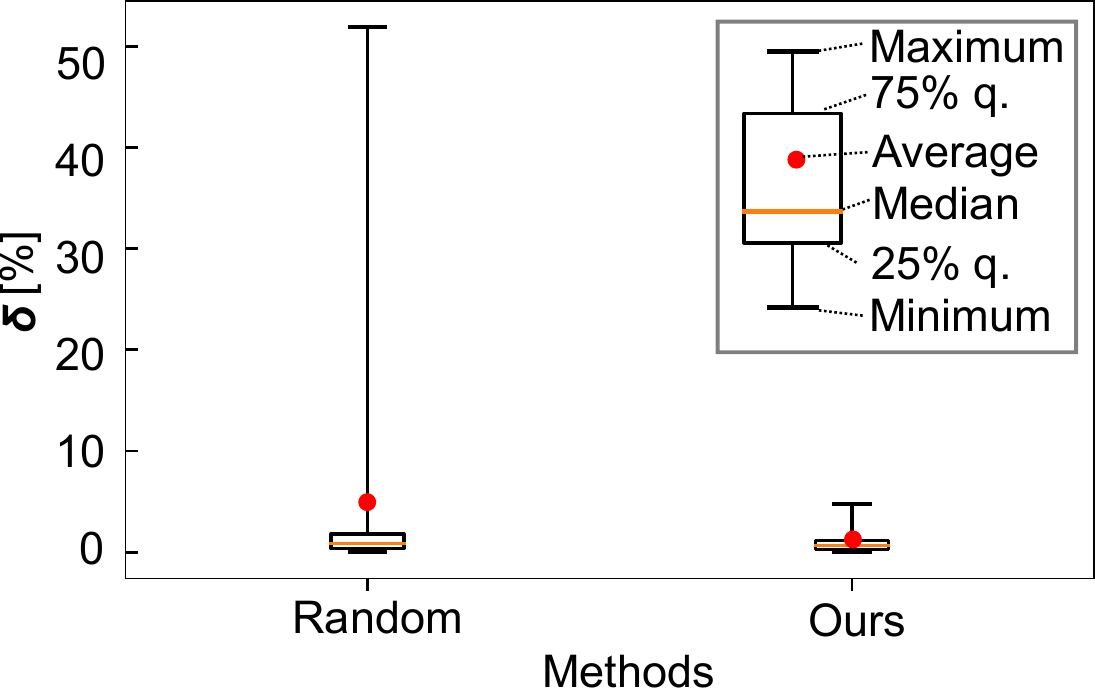}
  \caption{Box-plots of ${\bm \delta}$ for each method, where the
    maximum, 75\% quantile, median, average, 25\% quantile, and
    minimum values are shown. Because ${\bm \delta}$ of the random
    sampling is a multimodal distribution, the average is plotted out
    of the box.}
  \label{fig:p2_boxplot}
\end{figure}

\begin{figure}[t]
  \centering
  \hspace{-4mm}
  \subfigure[Random sampling]{
    \includegraphics[width=.6\linewidth]{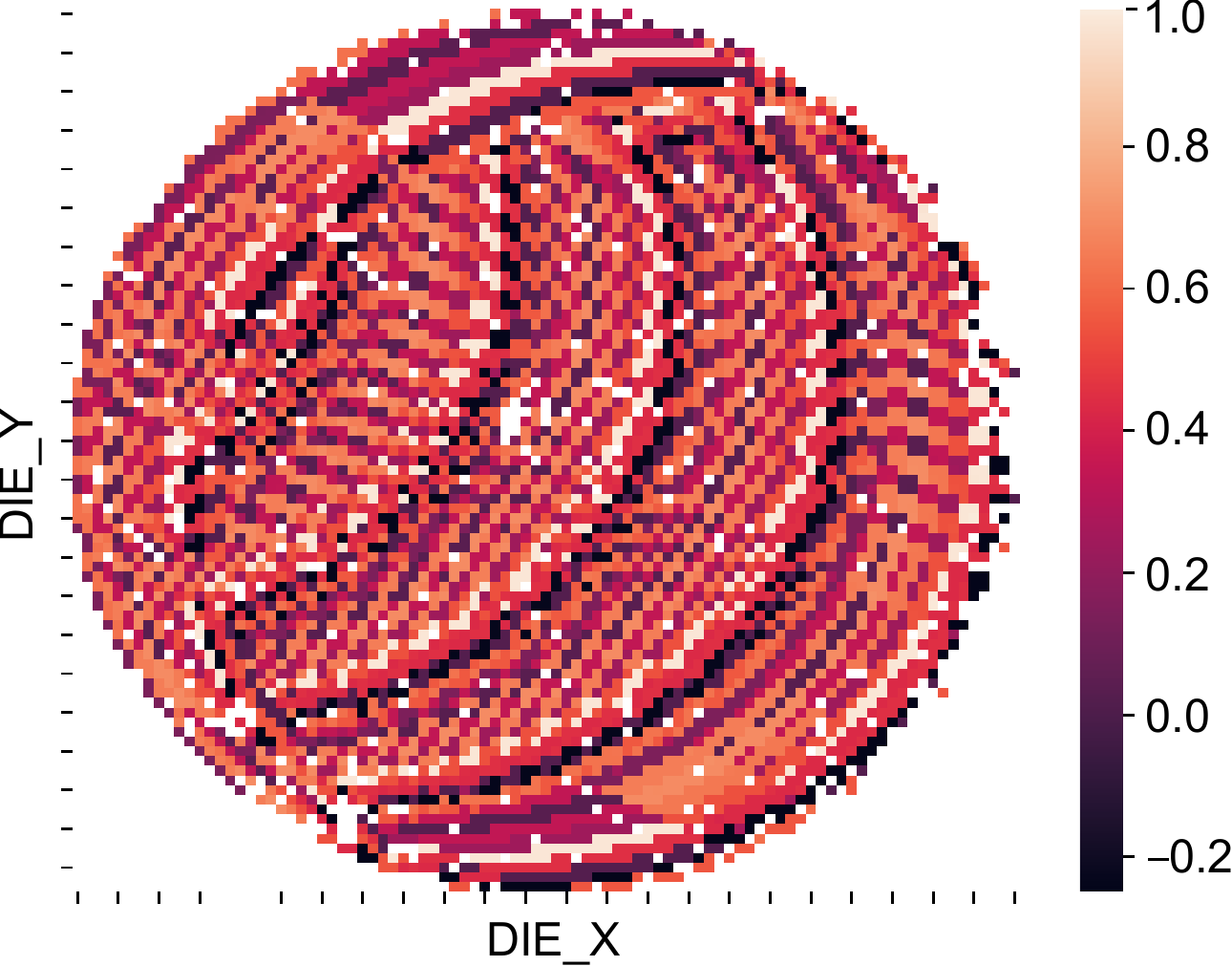}
    \label{fig:p2_random}
  }
  \subfigure[Ours]{
    \includegraphics[width=.57\linewidth]{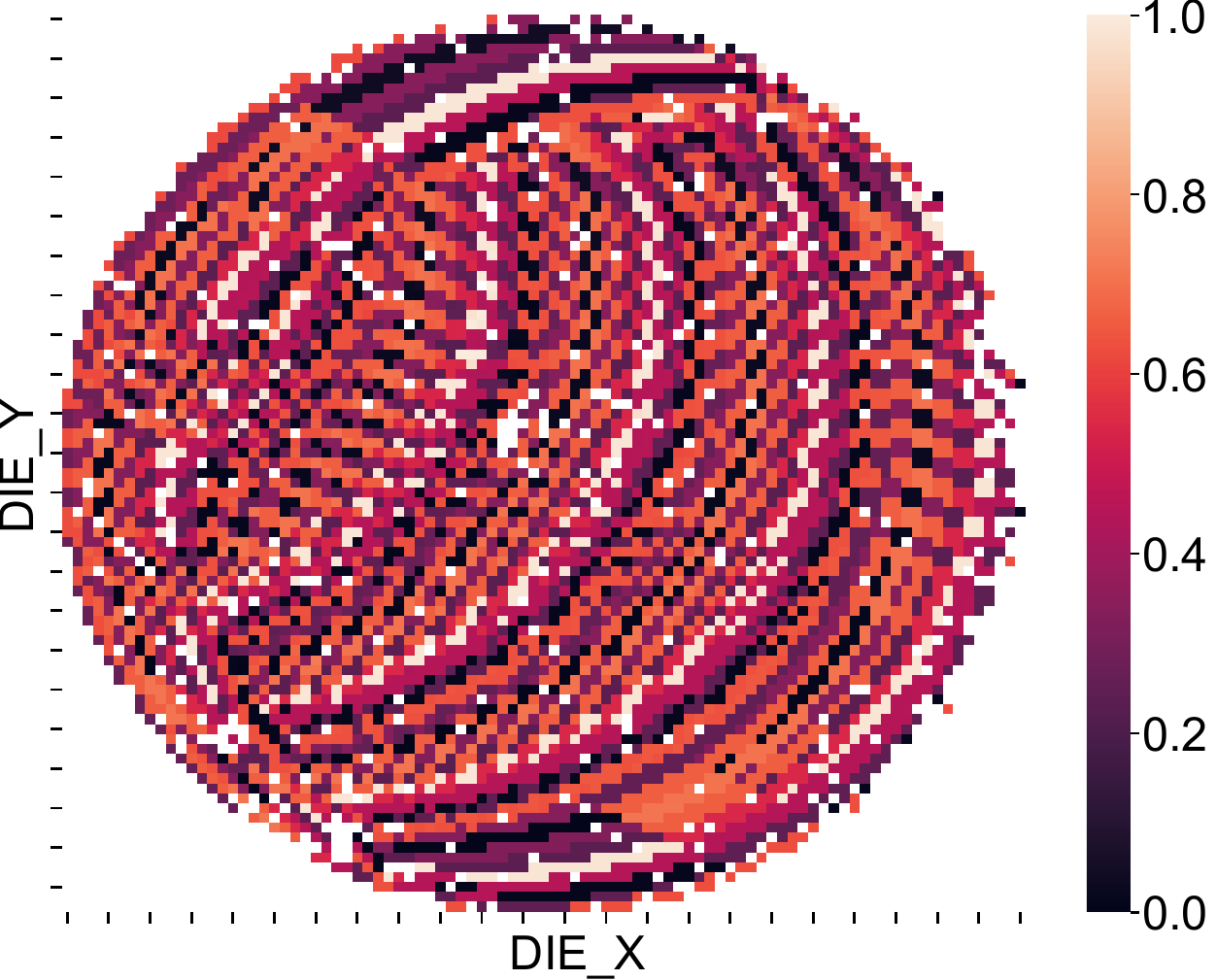}
    \label{fig:p2_ours}
  }
  \caption{Heat maps of the predicted characteristics for the 18
    touchdowns.}
  \label{fig:phase2}
\end{figure}

The box-plots for the random sampling and the proposed method for the
18 touchdowns are shown in Fig.~\ref{fig:p2_boxplot}. It is clear that
not only the average of ${\bm \delta}$ but also the variance of the
estimation errors can be reduced by the proposed method. Although the
random sampling has a maximum error of approximately 50\%, the
proposed method has only 4.77\%. In Fig.~\ref{fig:phase2}, the
prediction results for each method for the 18 touchdowns are
shown. Although they are visually similar, in
Fig.~\ref{fig:p2_random}, the predicted values exceed the range of the
normalized range, (i.e., 0 to 1), indicating that the predictions by
the random sampling are not sufficient. On the other hand, excellent
agreement is observed between Figs.~\ref{fig:p2_ours}
and~\ref{fig:full}.  It shows that the proposed method successfully
achieves highly accurate wafer-level variation modeling even in a
multi-site test environment.

\section{Conclusion}\label{sec:conclusion}
In this paper, we proposed a novel wafer-level spatial correlation
modeling method for multi-site RF IC testing. The proposed method
employs GP regression, which is an efficient statistical modeling
method used to predict the value for an unmeasured point from small
sampling data. In the proposed method, GP is applied individually by
partitioning the die location on a wafer according to the site
information provided by the test engineers. In addition, we propose an
active sampling method based on the predictive variance calculated by
GP to achieve better prediction results while maintaining a small
measurement cost. Experimental results using an industrial production
test dataset demonstrated that the proposed method achieves a 19.46
times smaller prediction error than the conventional method. Moreover,
we demonstrated that the proposed sampling method provides sufficient
prediction accuracy with 18 touchdown measurements, which corresponds
to 3\% of the number of touchdowns for full measurement. In contrast,
the prediction accuracy by random sampling required over 50
measurements. Although all the existing methods were evaluated with
one DUT measurement, the proposed method achieved better prediction
results by considering an actual touchdown under multi-site testing.

\bibliographystyle{IEEEtran}
\bibliography{alias,cad,ml}

\begin{thebibliography}{10}
\providecommand{\url}[1]{#1}
\csname url@samestyle\endcsname
\providecommand{\newblock}{\relax}
\providecommand{\bibinfo}[2]{#2}
\providecommand{\BIBentrySTDinterwordspacing}{\spaceskip=0pt\relax}
\providecommand{\BIBentryALTinterwordstretchfactor}{4}
\providecommand{\BIBentryALTinterwordspacing}{\spaceskip=\fontdimen2\font plus
\BIBentryALTinterwordstretchfactor\fontdimen3\font minus
  \fontdimen4\font\relax}
\providecommand{\BIBforeignlanguage}[2]{{%
\expandafter\ifx\csname l@#1\endcsname\relax
\typeout{** WARNING: IEEEtran.bst: No hyphenation pattern has been}%
\typeout{** loaded for the language `#1'. Using the pattern for}%
\typeout{** the default language instead.}%
\else
\language=\csname l@#1\endcsname
\fi
#2}}
\providecommand{\BIBdecl}{\relax}
\BIBdecl

\bibitem{TCAD2017_Wang}
L.-C. Wang, ``Experience of data analytics in {EDA} and test---principles,
  promises, and challenges,'' \emph{IEEE Transactions on Computer-Aided Design
  of Integrated Circuits and Systems}, vol.~36, no.~6, pp. 885--898, 2017.

\bibitem{ETS2018_Stratigopoulos}
H.-G. Stratigopoulos, ``Machine learning applications in {IC} testing,'' in
  \emph{Proceedings of IEEE European Test Symposium}, 2018.

\bibitem{ITC2018_Shintani}
M.~Shintani, M.~Inoue, and Y.~Nakamura, ``Artificial neural network based test
  escape screening using generative model,'' in \emph{Proceedings of IEEE
  International Test Conference}, 2018, p. 9.2.

\bibitem{TSM2010_Reda}
S.~Reda and S.~R. Nassif, ``Accurate spatial estimation and decomposition
  techniques for variability characterization,'' \emph{IEEE Transactions on
  Semiconductor Manufacturing}, vol.~23, no.~3, pp. 345--357, 2010.

\bibitem{ICCAD2009_Li}
X.~Li, R.~R. Rutenbar, and R.~D. Blanton, ``Virtual probe: {A} statistically
  optimal framework for minimum-cost silicon characterization of nanoscale
  integrated circuits,'' in \emph{Proceedings of IEEE/ACM International
  Conference on Computer-Aided Design}, 2009, pp. 433--440.

\bibitem{DAC2010_Zhang}
W.~Zhang, X.~Li, and R.~A. Rutenbar, ``Bayesian virtual probe: {M}inimizing
  variation characterization cost for nanoscale {IC} technologies via
  {Bayesian} inference,'' in \emph{Proceedings of ACM/EDAC/IEEE Design
  Automation Conference}, 2010, pp. 262--267.

\bibitem{TCAD2011_Zhang}
W.~Zhang, X.~Li, F.~Liu, E.~Acar, R.~A. Rutenbar, and R.~D. Blanton, ``Virtual
  probe: A statistical framework for low-cost silicon characterization of
  nanoscale integrated circuits,'' \emph{IEEE Transactions on Computer-Aided
  Design of Integrated Circuits and Systems}, vol.~30, no.~7, pp. 1814--1827,
  2011.

\bibitem{DATE2014_Zhang}
S.~Zhang, F.~Lin, C.-K. Hsu, K.-T. Cheng, and H.~Wang, ``Joint virtual probe:
  {J}oint exploration of multiple test items' spatial patterns for efficient
  silicon characterization and test prediction,'' in \emph{Proceedings of IEEE
  Design Automation and Test in Europe}, 2014.

\bibitem{ICCAD2012_Kupp}
N.~Kupp, K.~Huang, J.~M. Carulli, Jr., and Y.~Makris, ``Spatial correlation
  modeling for probe test cost reduction in {RF} devices,'' in
  \emph{Proceedings of IEEE/ACM International Conference on Computer-Aided
  Design}, 2012, pp. 23--29.

\bibitem{ITC2014_Ahmadi}
A.~Ahmadi, K.~Huang, S.~Natarajan, C.~John~M., Jr., and Y.~Makris,
  ``Spatio-temporal wafer-level correlation modeling with progressive sampling:
  A pathway to {HVM} yield estimation,'' in \emph{Proceedings of IEEE
  International Test Conference}, 2014, p. 18.1.

\bibitem{DATE2013_Huang}
K.~Huang, N.~Kupp, C.~John~M., Jr., and Y.~Makris, ``Handling discontinuous
  effects in modeling spatial correlation of wafer-level analog/{RF} tests,''
  in \emph{Proceedings of IEEE Design Automation and Test in Europe}, 2013, pp.
  553--558.

\bibitem{DATE2010_Marinissen}
E.~J. Marinissen, A.~Singh, D.~Glotter, M.~Esposito, J.~M.~C. {Jr.}, A.~Nahar,
  K.~M. Butler, D.~Appello, and C.~Portelli, ``Adapting to adaptive testing,''
  in \emph{Proceedings of IEEE Design Automation and Test in Europe}, 2010, pp.
  556--561.

\bibitem{ITC2011_Gotkhindikar}
K.~R. Gotkhindikar, W.~R. Daasch, K.~M. Butler, J.~M. Carulli, Jr., and
  A.~Nahar, ``Die-level adaptive test: Real-time test reordering and
  elimination,'' in \emph{Proceedings of IEEE International Test Conference},
  2011, p. 15.1.

\bibitem{ETS2012_Yilmaz}
E.~Yilmaz, S.~Ozev, O.~Sinanoglu, and P.~Maxwell, ``Adaptive testing:
  {Conquering} process variations,'' in \emph{Proceedings of IEEE European Test
  Symposium}, 2012.

\bibitem{TCAD2014_Shintani}
M.~Shintani, T.~Uezono, T.~Takahashi, K.~Hatayama, T.~Aikyo, K.~Masu, and
  T.~Sato, ``A variability-aware adaptive test flow for test quality
  improvement,'' \emph{IEEE Transactions on Computer-Aided Design of Integrated
  Circuits and Systems}, vol.~33, no.~7, pp. 1056--1066, 2014.

\bibitem{EM}
A.~P. Dempster, N.~M. Laird, and D.~B. Rubin, ``Maximum likelihood from
  incomplete data via the {EM} algorithm,'' \emph{Journal of the Royal
  Statistical Society. Series B (Methodological)}, vol.~39, no.~1, pp. 1--38,
  1977.

\bibitem{TIT2006_Donoho}
D.~L. Donoho, ``Compressed sensing,'' \emph{IEEE Transactions on Information
  Theory}, vol.~52, no.~4, pp. 1289--1306, 2006.

\bibitem{CS}
E.~J. Candes and M.~B. Wakin, ``An introduction to compressive sampling,''
  \emph{IEEE Signal Processing Magazine}, vol.~25, no.~2, pp. 21--30, 2008.

\bibitem{gp_book}
C.~E. Rasmussen and C.~K.~I. Williams, \emph{Gaussian Processes for Machine
  Learning}.\hskip 1em plus 0.5em minus 0.4em\relax MIT Press, 2006.

\bibitem{ITC2012_Sumikawa}
N.~Sumikawa, L.-C. Wang, and M.~S. Abadir, ``An experiment of burn-in time
  reduction based on parametric test analysis,'' in \emph{Proceedings of IEEE
  International Test Conference}, 2012, p. 19.3.

\bibitem{ETS2014_Lehner}
T.~Lehner, A.~Kuhr, M.~Wahl, and R.~Br\"{u}ck, ``Site dependencies in a
  multisite testing environment,'' in \emph{Proceedings of IEEE European Test
  Symposium}, 2014.

\bibitem{VTS2014_Farayola}
P.~O. Farayola, S.~K. Chaganti, A.~O. Obaidi, A.~Sheikh, S.~Ravi, and D.~Chen,
  ``Quantile -- quantile fitting approach to detect site to site variations in
  massive multi-site testing,'' in \emph{Proceedings of IEEE VLSI Test
  Symposium}, 2020.

\bibitem{BJMSP2006_Steinley}
D.~Steinley, ``{K-means clustering: A half-century synthesis},'' \emph{British
  Journal of Mathematical and Statistical Psychology}, vol.~59, no.~1, pp.
  1--34, 2006.

\bibitem{ITC2016_Butler}
K.~M. Butler, A.~Nahar, and W.~R. Daasch, ``What we know after twelve years
  developing and deploying test data analytics solutions,'' in
  \emph{Proceedings of IEEE International Test Conference}, 2016.

\bibitem{INNS-ENNS2000_Seo}
S.~Seo, M.~Wallat, T.~Graepel, and K.~Obermayer, ``Gaussian process regression:
  active data selection and test point rejection,'' in \emph{Proceedings of
  IEEE-INNS-ENNS International Joint Conference on Neural Networks}, 2000, pp.
  241--246.

\bibitem{prml}
C.~M. Bishop, \emph{Pattern Recognition and Machine Learning}.\hskip 1em plus
  0.5em minus 0.4em\relax Springer, 2006.

\bibitem{kernelcookbook}
D.~Duvenaud, ``The kernel cookbook,'' [Online]. Available:
  {https://www.cs.toronto.edu/~duvenaud/cookbook/}.

\bibitem{JMLR2001_Genton}
M.~G. Genton, ``Classes of kernels for machine learning: A statistics
  perspective,'' \emph{The Journal of Machine Learning Research}, vol.~2, pp.
  299--312, 2001.

\bibitem{TSM2004_Ohkawa}
S.~Ohkawa, M.~Aoki, and H.~Masuda, ``Analysis and characterization of device
  variations in an {LSI} chip using an integrated device matrix array,''
  \emph{IEEE Transactions on Semiconductor Manufacturing}, vol.~17, no.~2, pp.
  155--165, 2004.

\bibitem{TED2008_Saxena}
S.~Saxena, C.~Hess, H.~Karbasi, A.~Rossoni, S.~Tonello, P.~McNamara,
  S.~Lucherini, S.~Minehane, C.~Dolainsky, and M.~Quarantelli, ``Variation in
  transistor performance and leakage in nanometer-scale technologies,''
  \emph{IEEE Transactions on Electron Devices}, vol.~55, no.~1, pp.
  pp131--pp144, 2008.

\bibitem{JCAM1987_Rousseeuw}
P.~J. Rousseeuw, ``{Silhouettes: A} graphical aid to the interpretation and
  validation of cluster analysis,'' in \emph{Journal of Computational \&
  Applied Mathematics}, 1987, pp. 53--65.

\bibitem{NeuroImage1999_Goutte}
C.~Goutte, P.~Toft, E.~Rostrup, F.~A. Nielsen, and L.~K. Hansen, ``On
  clustering {fMRI} time series,'' \emph{NeuroImage}, vol.~9, pp. 298--310,
  1999.

\bibitem{CS1974_Calinski}
T.~Calinski and J.~Harabasz, ``A dendrite method for cluster analysis,''
  \emph{Communications in Statistics}, vol.~3, pp. 1--27, 1974.

\bibitem{JASA1991_Tang}
B.~Tang, ``Orthogonal array-based latin hypercubes,'' \emph{Journal of the
  American Statistical Association}, vol.~88, no. 424, pp. 1392--1397, 1991.

\bibitem{ACML2010_Park}
S.~Park and S.~Choi, ``Hierarchical {G}aussian process regression,'' in
  \emph{Proceedings of Asian Conference on Machine Learning}, 2010, pp.
  95--110.

\bibitem{SAC2019_Nguyen}
D.-T. Nguyen, M.~Filippone, and P.~Michiardi, ``Exact {G}aussian process
  regression with distributed computations,'' in \emph{Proceedings of
  ACM/SIGAPP Symposium on Applied Computing}, 2019, pp. 1286--1295.

\end{thebibliography}

\end{document}